\documentclass[journal]{IEEEtran}
\usepackage{bbm}
\usepackage{array}
\usepackage{dcolumn}
\usepackage{epsfig}
\usepackage[intlimits]{amsmath}
\usepackage{amsmath, amsfonts, yhmath, bm}
\usepackage{amssymb}
\usepackage{psfrag}
\usepackage{color,soul}
\usepackage[dvipsnames]{xcolor}
\usepackage[normalem]{ulem}
\usepackage{enumerate}
\usepackage{stackengine}
\usepackage[noadjust]{cite}
\usepackage{graphicx}
\usepackage{multirow}
\usepackage[caption=false,font=footnotesize]{subfig}
\usepackage{etoolbox}
\usepackage{tcolorbox}
\usepackage{float}
\usepackage{dsfont}
\usepackage{tikz}
\usetikzlibrary{arrows}
\usetikzlibrary{calc}
\usetikzlibrary{decorations.pathmorphing}
\usepackage{svg}
\usepackage{comment}
\usepackage{ifthen}

\usepackage{hyperref}
\hypersetup{
    colorlinks,
    linkcolor={red!50!black},
    citecolor={blue!50!black},
    urlcolor={blue!80!black}
}

\usepackage{balance}
\usepackage{url}
\usepackage[inline]{enumitem} 

\usetikzlibrary{backgrounds, positioning, fit}
\usepackage{pgfplots}
\pgfplotsset{compat=newest}  

\usepackage{mathtools} 

\usepackage{vmr-symbols-vecbold}
\usepackage{standard-macros}
\PassOptionsToPackage{hyphens}{url}

\newcommand{\defeq}{\mathrel{\mkern-0.25mu=}}
\DeclareSymbolFontAlphabet{\amsmathbb}{AMSb}%

\newcommand{\lro}[1]{\lefto({#1}\right)}																

\newcommand{\lr}[1]{\left({#1}\right)}																
\newcommand{\lrb}[1]{\left \lbrace {#1} \right \rbrace}															

\safemath{\dopplerspread}{B_D}																								
\safemath{\delayspread}{T_D}																									
\safemath{\nc}{n\sub{c}}																										
\safemath{\nf}{n\sub{f}}																										
\safemath{\efa}{p\sub{sc}}
\safemath{\efb}{p\sub{cs}}
\safemath{\ef}{\epsilon\sub{f}	}
\safemath{\nd}{n\sub{d}}																										
\safemath{\ntx}{n\sub{t}} 																											
\safemath{\nrx}{n\sub{r}}																											
\safemath{\ntxt}{\tilde{n\sub{t}}}																											
\safemath{\cb}{\ensuremath{L}} 																								
\safemath{\cl}{\ensuremath{n}} 																								
\safemath{\txanto}{{\ensuremath{\tilde{m}_t}}} 																		
\safemath{\cs}{M} 																														
\safemath{\idPustm}{\ensuremath{S_{k}}}
\safemath{\error}{\ensuremath{\epsilon}} 																				
\safemath{\eexp}{\ensuremath{\mathcal{E}}} 																			
\safemath{\nsubc}{n\sub{s}}			 																						
\safemath{\nofdm}{n\sub{o}} 																									
\safemath{\bc}{\ensuremath{B_c}} 																							
\safemath{\ts}{\ensuremath{T_s}} 																							
\safemath{\nrb}{\ensuremath{n_{rb}}} 																						
\safemath{\rul}{\ensuremath{\rho\sub{ul}}}
\safemath{\rdl}{\ensuremath{\rho\sub{dl}}}

\safemath{\nres}{\ell}
\safemath{\nr}{n\sub{r}}
\safemath{\maxk}{M^*\lr{\nres, \nsubc, \nofdm, \epsilon, \rho}}
\safemath{\Rmax}{R^*}
\safemath{\Emin}{E\sub{b}^*/N_0}
\safemath{\Eminf}{\frac{E\sub{b}^*}{N_0}}
\safemath{\np}{\ensuremath{n\sub{p}}}
\safemath{\ndf}{\ensuremath{\bar{n}\sub{d}}}
\safemath{\npf}{\ensuremath{\bar{n}\sub{p}}}
\safemath{\code}{\ensuremath{\mathcal{C}}}
\safemath{\err}{\ensuremath{\epsilon}}
\safemath{\rp}{\ensuremath{\rho\sub{p}}}
\safemath{\rd}{\ensuremath{\rho\sub{d}}}
\safemath{\cohtime}{\ensuremath{T\sub{c}}}
\safemath{\cohbw}{\ensuremath{B\sub{c}}}
\safemath{\nmax}{\ensuremath{\ell\sub{m}}}
\safemath{\ntot}{\ensuremath{n\sub{tot}}}
\safemath{\nul}{\ensuremath{n\sub{ul}}}
\safemath{\ndl}{\ensuremath{n\sub{dl}}}

\safemath{\yp}{\ensuremath{\randvecy_{\nu}^{(\text{p})}}}
\safemath{\yd}{\ensuremath{\randvecy_{\nu}^{(\text{d})}}}
\safemath{\ypd}{\ensuremath{\vecy_{\nu}^{(\text{p})}}}
\safemath{\ydd}{\ensuremath{\vecy_{\nu}^{(\text{d})}}}

\safemath{\ypf}{\ensuremath{\bar{\randvecy}_{\nu}^{(\text{p})}}}
\safemath{\ydf}{\ensuremath{\bar{\randvecy}_{\nu}^{(\text{d})}}}
\safemath{\ypdf}{\ensuremath{\bar{\vecy}_{\nu}^{(\text{p})}}}
\safemath{\yddf}{\ensuremath{\bar{\vecy}_{\nu}^{(\text{d})}}}

\safemath{\xp}{\ensuremath{\vecx^{(\text{p})}}}
\safemath{\xd}{\ensuremath{\randvecx_{\nu}^{(\text{d})}}}
\safemath{\xdd}{\ensuremath{\vecx_{\nu}^{(\text{d})}}}

\safemath{\xpf}{\ensuremath{\bar{\vecx}^{(\text{p})}}}
\safemath{\xdf}{\ensuremath{\bar{\randvecx}_{\nu}^{(\text{d})}}}
\safemath{\xddf}{\ensuremath{\bar{\vecx}_{\nu}^{(\text{d})}}}

\safemath{\xdb}{\ensuremath{\overline{\randvecx}^{(\text{d})}}}
\safemath{\Pxd}{\ensuremath{P_{\randvecx^{(\text{d})}}}}

\safemath{\xpbar}{\ensuremath{\overline{\matX}^{(\text{p})}}}
\safemath{\xdbar}{\ensuremath{\overline{\randmatX}^{(\text{d})}}}

\safemath{\xdv}{\ensuremath{\randvecx^{(\text{d})}}}
\safemath{\xdbarv}{\ensuremath{\overline{\randvecx}^{(\text{d})}}}
\safemath{\ydv}{\ensuremath{\randvecy^{(\text{d})}}}

\safemath{\xdr}{\ensuremath{\matX^{(\text{d})}}}

\safemath{\ttx}{\ensuremath{\tau\sub{tx}}}
\safemath{\trx}{\ensuremath{\tau\sub{rx}}}
\safemath{\ack}{\ensuremath{\mathrm{s}}}
\safemath{\nack}{\ensuremath{\mathrm{c}}}

\newcommand{\prob}[1]{\ensuremath{\mathbb{P}\lro{#1}}}

\safemath{\mI}{\ensuremath{i\lro{\randvecy ; \randvecx}}} 				

\safemath{\randveca}{\bm{A}}
\safemath{\randvecb}{\bm{B}}
\safemath{\randvecc}{\bm{C}}
\safemath{\randvecd}{\bm{D}}
\safemath{\randvece}{\bm{E}}
\safemath{\randvecf}{\bm{F}}
\safemath{\randvecg}{\bm{G}}
\safemath{\randvech}{\bm{H}}
\safemath{\randveci}{\bm{I}}
\safemath{\randvecj}{\bm{J}}
\safemath{\randveck}{\bm{K}}
\safemath{\randvecl}{\bm{L}}
\safemath{\randvecm}{\bm{M}}
\safemath{\randvecn}{\bm{N}}
\safemath{\randveco}{\bm{O}}
\safemath{\randvecp}{\bm{P}}
\safemath{\randvecq}{\bm{Q}}
\safemath{\randvecr}{\bm{R}}
\safemath{\randvecs}{\bm{S}}
\safemath{\randvect}{\bm{T}}
\safemath{\randvecu}{\bm{U}}
\safemath{\randvecv}{\bm{V}}
\safemath{\randvecw}{\bm{W}}
\safemath{\randvecx}{\bm{X}}
\safemath{\randvecy}{\bm{Y}}
\safemath{\randvecz}{\bm{Z}}
\safemath{\randvecphi}{\bm{\Phi}}

\safemath{\randmatA}{\amsmathbb{A}}
\safemath{\randmatB}{\amsmathbb{B}}
\safemath{\randmatC}{\amsmathbb{C}}
\safemath{\randmatD}{\amsmathbb{D}}
\safemath{\randmatE}{\amsmathbb{E}}
\safemath{\randmatF}{\amsmathbb{F}}
\safemath{\randmatG}{\amsmathbb{G}}
\safemath{\randmatH}{\amsmathbb{H}}
\safemath{\randmatI}{\amsmathbb{I}}
\safemath{\randmatJ}{\amsmathbb{J}}
\safemath{\randmatK}{\amsmathbb{K}}
\safemath{\randmatL}{\amsmathbb{L}}
\safemath{\randmatM}{\amsmathbb{M}}
\safemath{\randmatN}{\amsmathbb{N}}
\safemath{\randmatO}{\amsmathbb{O}}
\safemath{\randmatP}{\amsmathbb{P}}
\safemath{\randmatQ}{\amsmathbb{Q}}
\safemath{\randmatR}{\amsmathbb{R}}
\safemath{\randmatS}{\amsmathbb{S}}
\safemath{\randmatT}{\amsmathbb{T}}
\safemath{\randmatU}{\amsmathbb{U}}
\safemath{\randmatV}{\amsmathbb{V}}
\safemath{\randmatW}{\amsmathbb{W}}
\safemath{\randmatX}{\amsmathbb{X}}
\safemath{\randmatY}{\amsmathbb{Y}}
\safemath{\randmatZ}{\amsmathbb{Z}}
\safemath{\randmatSigma}{\mathbb{\Sigma}}
\safemath{\randmatPhi}{\mathbb{\Phi}}
\safemath{\randmatLambda}{\mathbb{\Lambda}}

\safemath{\matSigma}{\bm{\Sigma}}
\safemath{\matPhi}{\bm{\Phi}}
\safemath{\matLambda}{\bm{\Lambda}}

\newcommand{\midet}{\text{\tiny MID}}
\newcommand{\uc}{\text{\tiny UC}}
\newcommand{\lc}{\text{\tiny LC}}
\newcommand{\fa}{\text{\tiny FA}}
\newcommand{\md}{\text{\tiny MD}}
\newcommand{\detc}{\text{\tiny D}}

\newcommand{\mmse}{\text{\tiny MMSE}}

\newcommand{\eff}{\text{\tiny eff}}

\newcommand{\GLRT}{\text{\tiny GLRT}}

\newcommand{\indsub}{\text{\tiny ind}}
\newcommand{\quant}{\text{\tiny quant}}

\newcommand {\rectf} {\mathtt{rect}}

\newcommand {\bl} {\mathtt{BL}}
\newcommand{\ratedist}{\text{\tiny RD}}
\newcommand{\onebit}{\text{\tiny 1-bit}}
\newcommand{\fisherinf}{\text{\tiny FI}}

\definecolor{mitgrayback}{cmyk}{0.06, 0.05, 0.05, 0.0}
\definecolor{lightblack}{cmyk}{0.3, 0.3, 0.3, 0.25}
\definecolor{lightgray}{cmyk}{0.02, 0, 0, 0.0}
\definecolor{uppboundcolor}{HTML}{0072B2}

\newcommand{\addra}{}
\newcommand{\delra}[1]{}
\newcommand{\addraR}{}
\newcommand{\delraR}[1]{}

\makeatletter
\newsavebox\myboxA
\newsavebox\myboxB
\newlength\mylenA

\newcommand*\mybar[2][0.75]{%
	\sbox{\myboxA}{$\m@th#2$}%
	\setbox\myboxB\null
	\ht\myboxB=\ht\myboxA%
	\dp\myboxB=\dp\myboxA%
	\wd\myboxB=#1\wd\myboxA
	\sbox\myboxB{$\m@th\overline{\copy\myboxB}$}
	\setlength\mylenA{\the\wd\myboxA}
	\addtolength\mylenA{-\the\wd\myboxB}%
	\ifdim\wd\myboxB<\wd\myboxA%
	\rlap{\hskip 0.5\mylenA\usebox\myboxB}{\usebox\myboxA}%
	\else
	\hskip -0.5\mylenA\rlap{\usebox\myboxA}{\hskip 0.5\mylenA\usebox\myboxB}%
	\fi}
\makeatother

\makeatletter
\renewcommand*{\defeq}{\mathrel{\rlap{%
			\raisebox{0.3ex}{$\m@th\cdot$}}%
		\raisebox{-0.3ex}{$\m@th\cdot$}}%
	=}
\makeatother

\newtheorem{remark}{Remark}

\newtheorem{proposition}{Proposition}

\begin{document}

\title{Joint Extremum Compression and Detection of a Time-Delayed Signal for Distributed Sensing}

\author{Amir Weiss
\thanks{A. Weiss is with the Alexander Kofkin Faculty of Engineering, Bar-Ilan University, Ramat Gan, 5290002, Israel, e-mail: amir.weiss@biu.ac.il.}}

\maketitle

\begin{abstract}
We study the problem of joint compression and detection in distributed sensing systems, motivated by applications such as device-to-device connectivity in IoT networks and distributed radar. In such systems, spatially separated sensors must collaboratively decide whether their observations stem from a common underlying signal, while communicating over highly bandwidth-limited links. We consider a fundamental, insightful model in which one sensor (the encoder) observes a continuous-time realization of a stationary bandlimited Gaussian process, while the other sensor (the decoder) observes a delayed and noisy version of that signal, with an unknown delay. The encoder is allowed to transmit only a $k$-bit message to the decoder to assist in making a binary decision: either the observations are statistically independent, or they are time-shifted noisy versions of the same signal. We propose a low-complexity extremum-based scheme that exploits the structure of the signal to enable reliable decision-making under tight communication constraints. We derive nonasymptotic upper bounds on the false alarm and mis-detection probabilities of our method, as well as a simplified asymptotic bound for the latter. Representative simulations demonstrate that the proposed scheme outperforms the prevalent 1-bit-per-sample quantization baseline and\addra{ a Fisher-information-based compression benchmark, while closely approaching}\delra{ approaches} an information-theoretic (nonrealizable) rate--distortion benchmark.
\end{abstract}

\IEEEpeerreviewmaketitle
\vspace{-0.3cm}
\section{Introduction}\label{sec:introduction}
Detection is central to statistical inference in modern digital technology, enabling reliable decision-making in a wide range of systems including wireless communications~\cite{li2002mimo}, radar/sonar~\cite{diao2022review,abraham2002active}, biomedical monitoring~\cite{bachmann2012low} and cyber-physical systems~\cite{pasqualetti2013attack}, to name a few. In many such applications, the core task is to decide between competing possible models (e.g., signal-present vs.\ signal-absent) based on noisy observations, often under stringent system-level constraints, such as complexity, latency, bandwidth and reliability.

More specifically, \emph{distributed detection} has become increasingly important, driven by the proliferation of networked sensing platforms such as wireless sensor networks~\cite{chamberland2003decentralized,chen2006channelaware,puccinelli2005wireless}, Internet-of-Things (IoT) deployments~\cite{sarkar2014diat}, cooperative autonomous systems~\cite{qu2008cooperative}, and large-scale monitoring infrastructures~\cite{du2018sensable}. In these settings, spatially separated sensors observe related phenomena and must collectively infer a global hypothesis, typically via local processing followed by information fusion. Classical works and surveys have established fundamental architectures and performance limits for distributed detection and data fusion (e.g., \cite{tenney1981detection,chair1986optimal,viswanathan2002distributed,blum2002distributed,varshney1996distributed}).

A key bottleneck in distributed inference is \emph{communication}: sensors often cannot transmit raw observations due to limited bandwidth, power, or shared-medium contention. This creates an inherent tension between detection performance and the rate (or quality) of information exchanged. Consequently, the design of compression strategies tailored to detection has been studied under various---but mostly information-theoretic---formulations, including hypothesis testing with communication constraints and multiterminal compression \cite{ahlswede1986hypothesis,han1987hypothesis,shimokawa1994error}, lossy compression optimized for detection \cite{katz2016distributed}, noisy-channel variants \cite{sreekumar2018distributed,sreekumar2019distributed}, and more recent single-shot and multi-bit approaches \cite{carpi2021single,mao2024multi} alongside theoretical improvements~\cite{kochman2025improved}.
\begin{figure}[t]
            \centering
                             \begin{center}
    
    \tikzstyle{block} = [draw, fill=blue!20, rectangle, 
        minimum height=3em, minimum width=6em,node distance=2cm]
    \tikzstyle{smallblock} = [draw, fill=mitgrayback, rectangle, 
        minimum height=3em, minimum width=4em, node distance=2cm]
    \tikzstyle{reallysmallblockfixed} = [draw, fill=mitgrayback, rectangle, 
        minimum height=2em, minimum width=2em, node distance=1cm]
    \tikzstyle{reallysmallblockdesignable} = [draw, fill=blue!20, rectangle, 
        minimum height=2em, minimum width=2em, node distance=1.25cm]
    \tikzstyle{sum} = [draw, fill=mitgrayback, circle, node distance=3.5cm]
    \tikzstyle{source} = [coordinate]
    \tikzstyle{inputencoder} = [coordinate]
    \tikzstyle{noise} = [coordinate]
    \tikzstyle{output} = [coordinate, node distance=2.5cm]
    \tikzstyle{legendtext} = [coordinate]
    \hspace*{-0.25cm}
        \scalebox{0.45}{
        \begin{tikzpicture}[scale=0.85, mycirc/.style={circle,fill=teal, minimum size=0.1cm},auto, node distance=1.75cm,>=latex']
          \shade[inner color=teal, outer color=lightgray, opacity=0.4] (0,0) circle [radius=3+1];
          \begin{scope}[yshift=-0.2cm]
          \foreach \x in {0,1,2,3,4}
            {
              \ifthenelse{\NOT \x = 0}{
                \draw[PineGreen, opacity=0.5-1.2*0.\x,
                      decorate, decoration={snake, amplitude=1.5pt, segment length=20pt}]
                  (0,0) circle [radius=\x+1];
              }{
                \draw[PineGreen, opacity=0.75-0.\x,
                      decorate, decoration={snake, amplitude=1.5pt, segment length=20pt}]
                  (0,0) circle [radius=\x+1];
              }
            }
            \end{scope}

          \node[mycirc] (source) at (0,0) {};
          \node [above] at ($(source.north)+(0,0.8)$) {\LARGE{\color{teal}Source}};
          \node [below] at ($(source.north)-(0,5)$) {\LARGE{\color{teal}Exist under $\setH_1$}};
          \node [below] at ($(source.north)-(0,6)$) {\LARGE{Exist under $\setH_0 \cup \setH_1$ (always)}};
          \node [sum, right of=delay] at (3,2) (sum1) {$\Sigma$};
          \node [noise, above of=sum1] (noise1) {};
          \node[fit=(sum1) , draw, rectangle, rounded corners, line width=0.75pt, inner xsep=15pt, inner ysep=10pt, label=below:\LARGE{Sensor 1}] (frame1) {};
          \node [source, left of=sum1, xshift=0.2cm, yshift=0.55cm, label=below:{\LARGE{\color{teal}$\rnds(t)$}}] (sourcelabel1) {};
        
          \node [sum, right of=delay] at (4.5,-5) (sum2) {$\Sigma$};
          \node [noise, above of=sum2] (noise2) {};
          \node [output, right of=sum2] (output2) {};
          \node[fit=(sum2) , draw, rectangle, rounded corners, line width=0.75pt, inner xsep=15pt, inner ysep=10pt, label=below:\LARGE{Sensor 2}] (frame2) {};
          \node [source, left of=sum2, xshift=-0.3cm, yshift=0.2cm, label=below:{\LARGE{\color{teal}$\rnds(t-\delta)$}}] (sourcelabel2) {};
        
          \node [] at (14,2) (compressor) {\LARGE{Compressor}};
          \node[fit=(compressor) , draw, rectangle, rounded corners, line width=0.75pt, inner xsep=5pt, inner ysep=5pt] (frame3) {};
        
          \node [align=center, text depth=1ex] at (14,-1.5) (channel) {\LARGE{One-way $k$-bits} \\ \vspace{-0.25cm} \\ \LARGE{noiseless channel}};
          \node[fit=(channel) , draw, rectangle, rounded corners, line width=0.75pt, inner xsep=5pt, inner ysep=1pt] (frame4) {};
        
          \node [align=center, text depth=1ex] at (14,-5) (computingunit) {\LARGE{Detector}};
          \node[fit=(computingunit) , draw, rectangle, rounded corners, line width=0.75pt, inner xsep=7pt, inner ysep=7pt] (frame5) {};
        
          \draw [->, decorate, teal, decoration={snake, segment length=20pt, post length=2.5mm}, line width=0.8pt] (source) --  (sum1);
          \draw [->, line width=0.8pt] (noise1) -- node[yshift=0.3cm] {\LARGE{$\rndz_1(t)$}} (sum1);
          \draw [->, shorten >=5pt, line width=0.8pt] (sum1) -- node[xshift=0.2cm] {\LARGE{$\rndx(t)$}} (compressor);
          
          \draw [->, decorate, teal, decoration={snake, segment length=20pt, post length=1.5mm}, line width=0.8pt] (source) --  (sum2);
          \draw [->, line width=0.8pt] (noise2) -- node[yshift=0.4cm] {\LARGE{$\rndz_2(t)$}} (sum2);
          \draw [->, shorten >=7pt, line width=0.8pt] (sum2) -- node[xshift=0.1cm] {\LARGE{$\rndy(t)$}} (computingunit);
          \draw [->, shorten >=1pt, shorten <=5pt, line width=0.8pt] (compressor) -- (channel);
          \draw [->, shorten >=7pt, shorten <=1pt, line width=0.8pt] (channel) -- (computingunit);
          \node [output, right of=computingunit, xshift=1cm, line width=0.8pt] (estimator) {};
          \draw [->, shorten <=7pt, line width=0.8pt] (computingunit) -- node[yshift=0.05cm, xshift=0.1cm] {\LARGE{$\widehat{\setH}$}} (estimator);
        
          \node [fit=(frame2) (frame5) (sourcelabel2) (noise2), draw=gray, rectangle, rounded corners, line width=0.75pt, inner xsep=50pt, inner ysep=12pt, dashed, shift={(0.5cm,-0.3cm)}] (grouped) {};
          \node [fit=(frame1) (compressor) (sourcelabel1) (noise1), draw=gray, rectangle, rounded corners, line width=0.75pt, inner xsep=28pt, inner ysep=12pt, dashed, shift={(-0.1cm,-0.3cm)}] (grouped) {};
        
          \node [] at (15.7,3.7) (site1) {\LARGE{\color{gray}{Site 1}}};
          \node [] at (16,-3.3) (site2) {\LARGE{\color{gray}{Site 2}}};
        
        \end{tikzpicture}
        }

    \end{center}\vspace{-0.2cm}
    \caption{Problem setup and architecture. Under $\mathcal{H}_1$, observations are given by \eqref{eq:h1}; under $\mathcal{H}_0$ the source is absent and observations are given by \eqref{eq:h0}. Sensor~1 feeds the compressor and $k$ bits are sent over a noiseless one-way link to Site~2, where the detector fuses the message with $\rndy(t)$ to decide $\widehat{\setH}\in\{\setH_0,\setH_1\}$. All notations used in the figure are defined in Section~\ref{sec:problemformulation}.}
    \label{fig:illustration}\vspace{-0.5cm}
\end{figure}
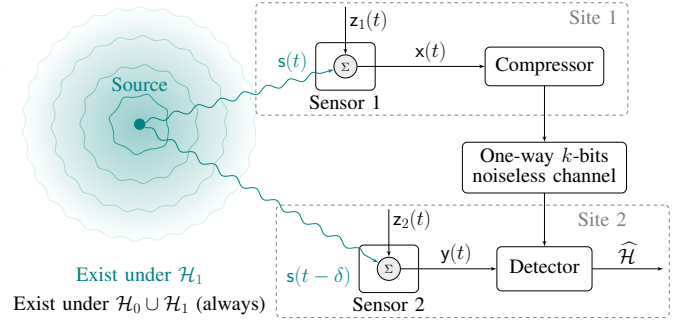

\vspace{-0.3cm}At the same time, a ubiquitous physical coupling between spatially separated sensors is \emph{propagation delay}: in many sensing modalities (e.g., acoustic, RF, seismic), signals arrive at different nodes with unknown relative time shifts. Such time-delay models, often termed as time-difference-of-arrival models, play a foundational role in detection/localization pipelines and have a long history in estimation and detection theory (e.g.,~\cite{chan1978least,weiss1983fundamentalpartI,weinstein1984fundamentalpartII,carter1987coherence}). However, despite the prevalence of delay-induced statistical structure in distributed systems, the literature on communication-constrained distributed detection has largely emphasized discrete-time memoryless models or generic dependence structures (e.g.,~\cite{escamilla2020distributed,inan2022fundamental,bounhar2024covert,bounhar2025dichotomy}), leaving a gap in understanding and designing rate-limited detectors that explicitly exploit (and are robust to) unknown time-delay coupling between sensors. This gap is particularly relevant in emerging distributed sensing applications where both (i) delay uncertainty is intrinsic and (ii) transmitting raw waveforms is bandwidth- and energy-intensive, and at times infeasible (e.g.,~\cite{fazel2012random,sani2016distributed,li2020data,musluoglu2025distributed}). Moreover, the now prevalent discrete-time model further assumes in many cases a discrete time-delay. This assumption is (almost) always an approximation, since the time-delay induced by the (arbitrary) underlying spatial configuration is rarely \emph{exactly} an integer multiplication of the digital system's sampling period. This approximation limits the attainable time resolution, and therefore leads to unnecessary estimation/detection performance degradation.

Motivated by this need for emerging sensing systems, we study distributed detection under a two-sensor time-delay propagation model with an explicit communication budget, as illustrated in Fig.~\ref{fig:illustration}. Building on the line of work~\cite{weiss2024joint,weiss2025extremum,weiss2025joint}, and further sharpening our preliminary work on a discrete-time approximated  model~\cite{weiss2025asilomar}, our contributions are threefold: 
\begin{enumerate}
    \item We formulate a more physically-faithful detection problem with continuous-time signals, and propose an appropriate realizable low-complexity encoder--detector (i.e., compression-detection) architecture that operates under a strict $k$-bit message constraint in a one-way link, while leveraging the continuous time-delay structure;
    \item For the proposed scheme, we provide a \emph{nonasymptotic} performance analysis, including bounds (and simplified approximations where helpful) for the false alarm (FA) and mis-detection (MD) probabilities that are \emph{explicit} in the underlying physical parameters, such as observation time, signal bandwidth and delay spread; and
    \item We corroborate our theoretical results and demonstrate the trade-off between communication and detection accuracy via simulations, comparing against both practical and theoretical (nonrealizable) representative baselines\addraR{, and representative robustness tests under multipath channels}.
\end{enumerate}

\addra{This paper should therefore be viewed as an achievability-type contribution: we formulate the continuous-time communication-constrained joint compression and detection problem, propose a realizable extremum-based scheme, and provide both nonasymptotic and asymptotic performance characterizations. A converse for this setting remains open and appears substantially more challenging than in standard rate--distortion problems, due to the combination of continuous-time observations, finite-rate encoding, and detection under an unknown continuous-valued time delay.}

The remainder of the paper is organized as follows. The following subsection introduces the notations used throughout the paper. In Section~\ref{sec:problemformulation} we formulate the observation model, hypotheses, and the communication-constrained distributed detection problem. Section~\ref{sec:proposedscheme} presents the proposed joint compression and detection scheme, and further provides its underlying intuition. Section~\ref{sec:performanceanalysis} develops the theoretical analysis and performance bounds. Section~\ref{sec:simulation} reports simulation results and comparisons to benchmarks, and Section~\ref{sec:conclusion} provides concluding remarks and an outlook for future work.

\vspace{-0.4cm}
\subsection{Notations}\label{subsec:notations}
We use lowercase letters with standard font and sans-serif font, e.g., $x$ and $\rndx$, to denote deterministic and random scalars, respectively. Similarly, we use $\vecx$ and $\rvecx$ for deterministic and random vectors, respectively; and $\matX$ and $\rmatX$ for deterministic and random matrices, respectively. Calligraphic letters (e.g., $\setA,\setX$) denote sets, and blackboard-bold letters (e.g., $\reals$) denote standard number spaces. The probability of an event $\setA$ is denoted by $\Prob(\setA)$, and the expectation and variance of a random variable (RV) $\rndx$ are denoted by $\Exop[\rndx]$ and $\Varop(\rndx)$, respectively. We use $\tp{(\cdot)}$ and $\herm{(\cdot)}$ to denote transpose and Hermitian transpose, respectively. The indicator function $\mathbbm{1}_{\setA}$ is equal to one if $\setA$ is true and zero otherwise. The standard floor and sign operators are denoted by $\left\lfloor \cdot \right\rfloor$ and $\sign(\cdot)$, respectively. We write $f=o(g)$ to indicate that $\lim f/g=0$, where the arguments and implied limits should be clear from the context.

\section{Problem Formulation}\label{sec:problemformulation}
We study joint compression and detection in a distributed sensing setting, where two spatially separated sensors observe continuous-time signals and must decide whether they are independent or time-shifted versions of a common source. Communication is limited to a unidirectional $k$-bit message from the first sensor (referred to as ``encoder") to the second (referred to as ``decoder"), which then makes the decision.

Let $\rndx(t)$ and $\rndy(t)$ denote the continuous-time signals observed by the encoder and the decoder, respectively, over an interval $t \in (-\delta, T+\delta)$. The detection task consists in deciding between the following two hypotheses:
\begin{align}
    \setH_0:\; &\begin{cases}
        \rndx(t) = \rndz_1(t), \\
        \rndy(t) = \rndz_2(t),
    \end{cases}
    \quad &&\text{(no signal)} \label{eq:h0} \\
    \setH_1:\; &\begin{cases}
        \rndx(t) = \rnds(t) + \rndz_1(t), \\
        \rndy(t) = \rnds(t - \delta) + \rndz_2(t),
    \end{cases}
    \quad &&\text{(signal present)} \label{eq:h1}
\end{align}
where:
\begin{itemize}
    \item The noise signals $\rndz_1(t)$ and $\rndz_2(t)$ are independent, zero-mean, stationary Gaussian processes with variances $\sigma_1^2$ and $\sigma_2^2$, respectively, and power spectral densities (PSDs)
    \begin{equation}\label{eq:psdnoise}
        S_{\rndz_i\rndz_i}(f) = \frac{\sigma_i^2}{B} \rectf\left(\frac{f}{B}\right), \quad i = 1,2,
    \end{equation}
    where $ \rectf(x) \triangleq \mathbbm{1}_{|x|<\frac{1}{2}} $ is the standard rectangular function, $\sigma_1,\sigma_2\in\positivereals$ are considered as deterministic unknowns parameters, and $B \in \positivereals$ is the bandwidth;
    
    \item The common signal $\rnds(t)$ is a zero-mean stationary Gaussian process, bandlimited to bandwidth $B$, independent of $\rndz_1(t)$ and $\rndz_2(t)$, with an autocorrelation function
    \begin{equation}
        R_{\rnds\rnds}(\tau) \triangleq \Exop[\rnds(t)\rnds(t - \tau)],
    \end{equation}
    variance $\Varop(\rnds(t)) = R_{\rnds\rnds}(0) = \sigma_{\rnds}^2$, and PSD
    \begin{equation}
        S_{\rnds\rnds}(f) = \setF\left\{ R_{\rnds\rnds}(\tau) \right\} = \frac{\sigma_{\rnds}^2}{B} \rectf\left( \frac{f}{B} \right);
    \end{equation}
    
    \item $\delta \in \Delta \triangleq [-\delta_m, \delta_m]$ is an unknown deterministic relative time delay, where $|\Delta|=2\delta_m$ is the delay spread, and we assume $1/B<2\delta_m$.\footnote{The assumption $1/B<2\delta_m$ means that the bandwidth of the signal $\rnds(t)$ allows for a sufficiently high time-resolution, and specifically that the sampling period under Nyquist sampling, $T_{\mathrm{s}}=1/B$, is smaller than $2\delta_m$.}
\end{itemize}
It follows that the conditional cross-correlation between the signals $\rndy(t)$ and $\rndx(t)$ given $\setH$ satisfies
\begin{equation}
    \Exop\left[ \rndy(t)\rndx(t - \tau) \mid \setH \right] = 
    \begin{cases}
        0, & \setH = \setH_0 \\
        R_{\rnds\rnds}(\tau - \delta), & \setH = \setH_1
    \end{cases}.
    \label{eq:crosscorr}
\end{equation}

The encoder observes $\rndx(t)$ over $t \in (0, T)$ and transmits a $k$-bit message $\rvecm \in \{0,1\}^{k\times 1}$ to the decoder over a noiseless channel. The decoder, which observes $\rndy(t)$, applies a decision rule
\begin{equation}\label{eq:totalerrprob}
    \widehat{\setH}: \{0,1\}^{k} \times \bl \to \{\setH_0, \setH_1\},
\end{equation}
where $\bl$ denotes the space of wide-sense stationary (WSS), bandlimited processes.

The goal is to minimize the MD probability (declaring $\setH_0$ given $\setH=\setH_1$) for a fixed tolerance on the FA probability (declaring $\setH_1$ given $\setH=\setH_0$). Formally, for a FA tolerance level $\epsilon_{\fa} \in [0,1]$, we seek for the best \emph{joint} compression and detection scheme with respect to the optimality criterion
\begin{align}\label{eq:optimizationproblem}
    &\inf_{\substack{f: \bl(0,T) \, \to \, \{0,1\}^{k\times 1} \\ \widehat{\setH}: \{0,1\}^{k\times 1} \times \bl \to \{\setH_0, \setH_1\}}} &&\prob{\widehat{\setH}(\rvecm, \rndy(t)) = \setH_0 \mid \setH_1 },\\
    &\quad\quad\quad\quad\,\;\mathrm{s.t.} &&\prob{\widehat{\setH}(\rvecm, \rndy(t)) = \setH_1 \mid \setH_0 }\leq \epsilon_{\fa},
\end{align}
where $f$ is the encoding mapping from the space of WSS bandlimited processes observed over the time interval $(0,T)$ to the set of all $k$-bit messages, and the probability is taken over the signal and noise processes. For simplicity, the hypothesis prior is assumed to be uniform, i.e., $\prob{\setH_0}=\prob{\setH_1}=\frac{1}{2}$.

We note that the modeling choice of a $k$-bit message over a noiseless channel, e.g., as in~\cite{zhang1988estimation,hadar2019distributed,kochman2021communication,cai2024distributed}, is an abstraction used mainly to isolate the \emph{fundamental impact of communication constraints on inference performance}: it captures a strict information-flow limitation while avoiding ancillary effects due to, for example, coding, retransmissions, or link outages. In practice, a sufficiently reliable physical link with appropriate channel coding can operate at an arbitrarily small error probability\addra{ only under idealized asymptotic conditions, e.g., sufficiently large blocklength and the ability to tolerate the associated latency. Thus, the present abstraction is most appropriate in regimes where link-layer reliability can be made high enough that}\delra{, so} the dominant bottleneck \delra{becomes}\addra{is} the available bit budget, which directly dictates the design and performance of the compression-detection architecture.\addra{ In finite-blocklength or channel-uncertain settings, additional communication impairments may also affect performance, but accounting for such effects is beyond the scope of this particular model.}

\vspace{-0.1cm}
\section{Proposed Joint Compression-Detection Scheme}\label{sec:proposedscheme}
Following recent work on extremum-based time-delay estimation\addra{ (TDE)} in distributed sensing~\cite{weiss2024joint,weiss2025extremum,weiss2025joint}, we propose a low-complexity scheme that enables effective detection under stringent communication constraints, as follows.

\vspace{0.1cm}
\noindent\textbf{Sampling and Extremum Encoding:} The encoder samples the continuous-time signal $\rndx(t)$ at its Nyquist rate, i.e., at a sampling period $T_{\mathrm{s}} = 1/B$, to obtain the discrete sequence
\begin{equation}
    \rndx[n] \triangleq \left.\rndx(t)\right|_{t=nT_{\mathrm{s}}} \quad \Longrightarrow \quad \setX_N \triangleq \{\rndx[n]\}_{n=0}^{N-1},
\end{equation}
where we set $N = 2^k$ such that $T=NT_{\mathrm{s}}$. The encoder transmits the index of the maximum sample in $\setX_N$:
\begin{equation}
    \rndj \triangleq \arg \max_{0 \leq n < N} \; \rndx[n],
    \label{eq:encoder_rule}
\end{equation}
which can be represented using exactly $k = \log_2 N$ bits. Thus, $\rvecm\in \{0,1\}^{k}$ is set as the binary representation of $\rndj$.

\vspace{0.1cm}
\noindent\textbf{Maximum-Index-Based Detection:} The decoder receives $\rvecm$, namely the index $\rndj$, and recovers the corresponding time
\begin{equation}
t_{\rndj} \triangleq \rndj T_{\mathrm{s}}\in T_{\mathrm{s}}\naturals\subset\reals_+\cup\{0\}.    
\end{equation}
It then evaluates the continuous-time signal $\rndy(t)$ within the delay uncertainty interval $\Delta$, shifted relative to $t_{\rndj}$. The test statistic is defined as
\begin{equation}
    \rndt_{\midet} \triangleq \sup_{\tau \in \Delta} \; \rndy(t_{\rndj} + \tau),
    \label{eq:test_stat_ct}
\end{equation}
and the decoder declares $\setH_1$ if this statistic exceeds a pre-defined threshold $\gamma \in \reals$. Formally, the decision rule is:
\begin{equation}
    \widehat{\setH}(\rvecm, \rndy(t)) =
    \begin{cases}
        \setH_1, & \rndt_{\midet} \geq \gamma, \\
        \setH_0, & \rndt_{\midet} < \gamma.
    \end{cases}
    \label{eq:detector_rule_ct}
\end{equation}

The decoder treats $t_{\rndj}$ as an ``anchor'' in time, \delra{at which the highest signal energy was observed}\addra{corresponding to the largest observed sample} at the encoder side under $\setH_1$. The decoder then scans its received signal $\rndy(t)$ in a $\Delta$ neighborhood around this point to detect a potential signal match. Under $\setH_1$, when $\rndy(t)$ aligns with the anchor, it produces a noticeable peak. Under $\setH_0$, the signal is pure noise, and the maximum behaves according to the noise distribution. As we shall demonstrate both analytically and empirically, this difference enables reliable detection even under severe communication constraints, conveying only a handful of bits.

\addra{It is important to note that, under the Nyquist-sampled Gaussian model, the transmitted index $\rndj$ should not be interpreted as estimating a latent physically meaningful location in the waveform. Since the samples $\{\rndx[n]\}_{n=0}^{N-1}$ are i.i.d., the maximizer index $\rndj=\arg\max_n \rndx[n]$ is uniform on $\{0,\ldots,N-1\}$ by symmetry. Its utility comes instead from the fact that conditioning on $\rndj$ selects an atypically large sample value $\rndx[\rndj]$, which under $\setH_1$ induces a locally more informative aligned decoder signal than the surrounding non-aligned alternatives.}

\addra{We note that in the implementation of the proposed scheme, the $k$-bit budget and the observation window are linked through $N=2^k$, so that for a fixed bandwidth $B$, one has $T = 2^k/B$. This explicit relation implies that increasing $k$ in this architecture also increases the required sensing duration, which may be restrictive in latency-limited regimes.}

\subsection{An Intuitive Interpretation of the Proposed Scheme via the GLRT with Unlimited Communication}
\label{subsec:glrt_unlimited}

We now consider an \emph{ideal} benchmark in which there is no communication
constraint between the sensors. A fusion center has full access to the
continuous-time signals $\rndx(t)$ and $\rndy(t)$ over $t\in(0,T)$ and must
decide between $\setH_0$ and $\setH_1$.

In Appendix~\ref{app:GLRT} we derive the corresponding GLRT in a continuous-time setting. The derivation proceeds by projecting the pair $(\rndx(t),\rndy(t+\tau))$ onto an arbitrary orthonormal basis of $L^2(0,T)$, applying the finite-dimensional Gaussian GLRT on the resulting coefficients, and then letting the projection dimension grow. Using Parseval’s identity applied pathwise to the random signals, we show that for each $\tau\in\Delta$ the (maximized) log-likelihood ratio is a strictly increasing function of the normalized empirical cross-correlation
\begin{equation}
    \widehat{\rho}_{\uc}(\tau)
    \triangleq
    \frac{\displaystyle\int_0^T \rndx(t)\rndy(t+\tau)\,\mathrm{d}t}{
    \sqrt{\displaystyle\int_0^T \rndx^2(t)\,\mathrm{d}t
         \displaystyle\int_0^T \rndy^2(t+\tau)\,\mathrm{d}t}}.\label{eq:normalizedcrosscorr}
\end{equation}
Consequently, the GLRT in the unlimited-communication setting can be written as
the correlation test
\begin{equation}
    \max_{\tau\in\Delta} \widehat{\rho}_{\uc}(\tau)
    \;\mathop{\gtrless}_{\setH_0}^{\setH_1}\; \gamma_{\uc},
    \label{eq:ct_glrt_corr}
\end{equation}
for a threshold $\gamma_{\uc}\in\mathbb{R}$ chosen to satisfy a FA tolerance.

Interestingly, our proposed extremum-based scheme shares the same core structure as this GLRT: both perform a delay-maximization test on a normalized empirical cross-correlation functional, differing only in how the correlation is estimated under the communication constraint\delra{, as we show next}.\addra{ The key issue is then to preserve \emph{where} to align the two observations in the presence of an unknown delay. In this sense, the extremum index is particularly valuable, since it conveys timing information that localizes the relevant neighborhood in $\rndy(t)$ using only $k$ bits. By contrast, amplitude- or energy-only summaries may reflect signal strength, but do not directly identify the alignment location. Moreover, in the Gaussian model considered here, the amplitude of the maximum is asymptotically less informative, since classical extreme-value results~\cite{HandbookExtremes2026} imply that it converges to deterministic value as $k$ grows.}

In Appendix~\ref{app:unbiasedness}, we further show that when the variances $\sigma_{1}^2, \sigma_{2}^2$ and $\sigma_{\rnds}^2$ are known, the extremum-based function
\begin{equation}\label{eq:ccdefcommconst}
    \widehat{\rho}_{\lc}(\tau)
    \triangleq \frac{\sqrt{\sigma_{\rnds}^2+\sigma_1^2}}{\sqrt{\sigma_{\rnds}^2+\sigma_2^2}}\cdot \frac{\rndy(t_{\rndj} + \tau)}{\Exop\left[\rndx[\rndj] \mid \setH_1\right]}
\end{equation}
is an unbiased estimator of the correlation function between $\rndy(t)$ and $\rndx(t)$, while relying only on the $k$-bit index of the encoder extremum and on $\rndy(t)$. That is,
\begin{equation}\label{eq:normcorrestunbiased}
    \Exop\left[\widehat{\rho}_{\lc}(\tau)\right] = \frac{\Exop\left[ \rndy(t)\rndx(t-\tau) \mid \setH \right]}{\sqrt{\Exop\left[\rndy^2(t) \mid \setH \right]\Exop\left[ \rndx^2(t) \mid \setH \right]}} \triangleq \rho(\tau).
\end{equation}
In particular, using the same extremum time $t_{\rndj}$ that underlies the message $\rvecm$, the decoder can evaluate $\widehat{\rho}_{\lc}(\tau)$---and, when $\sigma_{1}^2, \sigma_{2}^2$ and $\sigma_{\rnds}^2$ are unknown, a scaled version of $\widehat{\rho}_{\lc}(\tau)$---for all $\tau\in\Delta$ without any additional information from the encoder.

Rewriting the test statistic~\eqref{eq:test_stat_ct} using
\eqref{eq:ccdefcommconst}, we obtain
\begin{align}
    \rndt_{\midet}
    &= \sup_{\tau\in\Delta} \rndy(t_{\rndj} + \tau) \\
    &= \sup_{\tau\in\Delta}
       \frac{\sqrt{\sigma_{\rnds}^2+\sigma_2^2}}{\sqrt{\sigma_{\rnds}^2+\sigma_1^2}}\Exop\left[\rndx[\rndj] \mid \setH_1\right]\cdot\widehat{\rho}_{\lc}(\tau) \\
    &= \frac{\sqrt{\sigma_{\rnds}^2+\sigma_2^2}}{\sqrt{\sigma_{\rnds}^2+\sigma_1^2}}\Exop\left[\rndx[\rndj] \mid \setH_1\right]\cdot \sup_{\tau\in\Delta} \widehat{\rho}_{\lc}(\tau).
    \label{eq:test_stat_ct_rho}
\end{align}
Since the scalar factor
$\Exop\left[\rndx[\rndj] \mid \setH_1\right]\cdot\sqrt{\sigma_{\rnds}^2+\sigma_2^2}/\sqrt{\sigma_{\rnds}^2+\sigma_1^2}$ does not depend on $\tau$, it has no effect on the threshold test in~\eqref{eq:detector_rule_ct} other than a constant scaling. Hence, our detector in the limited-communication setting can be expressed as 
\begin{equation}
    \max_{\tau\in\Delta} \widehat{\rho}_{\lc}(\tau)
    \;\mathop{\gtrless}_{\setH_0}^{\setH_1}\; \widetilde{\gamma}_{\lc},
    \label{eq:ct_extremum_corr}
\end{equation}
where $\widetilde{\gamma}_{\lc} \triangleq \gamma \cdot \sqrt{\sigma_{\rnds}^2+\sigma_1^2} / \left(\Exop\left[\rndx[\rndj] \mid \setH_1\right]\cdot\sqrt{\sigma_{\rnds}^2+\sigma_2^2} \right)$, and can therefore be interpreted as \emph{thresholding the maximum, over $\tau\in\Delta$, of an extremum-based normalized empirical cross-correlation estimate}, in direct analogy with the GLRT~\eqref{eq:ct_glrt_corr}, which thresholds the maximum of the normalized empirical cross-correlation~\eqref{eq:normalizedcrosscorr} with unlimited communication.

Thus, the proposed extremum-based scheme \eqref{eq:ct_extremum_corr} preserves the core ``maximized correlation'' structure of the GLRT \eqref{eq:ct_glrt_corr}, but replaces the estimator \eqref{eq:normalizedcrosscorr} with a $k$-bit extremum-based estimator. This connection formalizes the intuition that our detector is effectively mimicking the operation of the classical correlation test, but under tight communication constraint.

\section{Performance Analysis}\label{sec:performanceanalysis}

We now analyze the detection performance of the proposed scheme in terms of its FA and MD probabilities. Throughout, $\gamma\in\positivereals$ denotes a prespecified threshold for the test \eqref{eq:detector_rule_ct}. The propositions below, whose proofs appear in Sections \ref{subsec:falsealaramprob} and \ref{subsec:misdetectionprob}, respectively, provide nonasymptotic upper bounds on the FA and MD probabilities.

\begin{proposition}[FA probability upper bound]\label{prop1}
Denote the probability of a FA event as
\begin{equation}
    P_{\fa}(\gamma) \triangleq \prob{ \sup_{\tau \in \Delta} \rndy(t_{\rndj} + \tau) > \gamma \mid \setH_0 }.
\end{equation}
Then,
\tcbset{colframe=gray!90!blue,size=small,width=0.49\textwidth,halign=flush center,arc=2mm,outer arc=1mm}
\begin{tcolorbox}[upperbox=visible,colback=white,halign=left]
    \begin{equation}
    P_{\fa}(\gamma) \leq Q\left( \frac{\gamma}{\sigma_2} \right)
    + \frac{\delta_m B}{\sqrt{3}} \cdot \exp\left( -\frac{\gamma^2}{2\sigma_2^2} \right),\label{eq:falsealaramupperbound}
    \end{equation}
\end{tcolorbox}
\noindent where $Q(x)\triangleq \int_{x}^{\infty}\frac{1}{\sqrt{2\pi}}e^{-\frac{t^2}{2}}{\rm d}t$ is the standard $Q$-function.
\end{proposition}

\begin{remark}
One may obtain the slightly looser, though simpler upper bound,
\begin{equation}
    P_{\fa}(\gamma) < 
    \left( \frac{\sigma_2}{\gamma\sqrt{2\pi}} + \frac{\delta_m B}{\sqrt{3}} \right)\cdot \exp\left( -\frac{\gamma^2}{2\sigma_2^2} \right),
\end{equation}
by using $Q(x)< \frac{1}{x\sqrt{2\pi}}e^{-\frac{x^2}{2}}$, which holds for any $x>0$.\addra{ In addition, \eqref{eq:falsealaramupperbound} makes the parameter dependence transparent: for fixed $\delta_m$ and $B$, increasing the threshold $\gamma$ suppresses the FA probability exponentially. Also, while both $\sigma_2$ and $\delta_m B$ adversely affect FA performance, their effects are of different orders: the impact of $\sigma_2$ is more severe, since it appears in the exponent, whereas $\delta_m B$ enters only as a linear prefactor, reflecting the width of the searched noise-only region.}
\end{remark}

\begin{proposition}[MD probability upper bound]\label{prop2}
Denote the probability of a MD event as
\begin{equation}
    P_{\md}(\gamma) \triangleq \prob{ \sup_{\tau \in \Delta} \rndy(t_{\rndj} + \tau) < \gamma \mid \mathcal{H}_1 }.
\end{equation}
Let $\Phi_2(t_1,t_2;\rho)$ denote the standard bivariate normal cumulative distribution function (CDF) with correlation $\rho$, i.e.,
\begin{equation}
    \Phi_2(t_1,t_2;\rho)
    \triangleq \int_{-\infty}^{t_1}\phi(v)\,\Phi\!\left(\frac{t_2-\rho v}{\sqrt{1-\rho^2}}\right)\,{\rm d}v,
\end{equation}
where $\phi(t)\triangleq \frac{1}{\sqrt{2\pi}}e^{-\frac{t^2}{2}}$, and let $\Phi\left(\cdot\right)$ and $\Phi^{-1}\left(\cdot\right)$ denote the standard normal CDF and its inverse, respectively. Define the ancillary (non-negative) parameters
\begin{align}
\sigma_{\rndx}^2 &\triangleq \sigma_{\rnds}^2 + \sigma_{1}^2, \qquad\quad\; \beta \triangleq \frac{\sigma^2_{\rnds}}{\sigma^2_{\rndx}},\label{eq:defofmmseconstants}\\
\sigma_{\mmse}^2 &\triangleq \sigma_2^2 + \beta\sigma_1^2, \qquad\, \sigma_{\eff}^2 \triangleq \beta^2\sigma_{\rndx}^2 + \sigma_{\mmse}^2,\label{eq:defofmmseconstants2}
\end{align}
and the (non-negative) counting functions
\begin{align}
M_{\mathrm{in}}(j) &\triangleq \min\left\{L,\,j\right\} + \min\left\{L,\,N-1-j\right\},\label{eq:minhowmuch} \\
M_{\mathrm{out}}(j) &\triangleq 2L - M_{\mathrm{in}}(j),\label{eq:mouthowmuch}
\end{align}
where
\begin{equation}\label{eq:defofintL}
L\triangleq \left\lfloor \tfrac{\delta_m}{T_{\mathrm{s}}} \right\rfloor = \left\lfloor \delta_m B \right\rfloor \in\naturals\cup\{0\},    
\end{equation}
with which we further define
\begin{align}\label{eq:defofwhat}
        B_{\mathrm{in}}(x,\gamma) &\triangleq \frac{\Phi_2\!\left(\tfrac{x}{\sigma_{\rndx}},\tfrac{\gamma}{\sigma_{\eff}};\frac{\beta\sigma_{\rndx}}{\sigma_{\eff}}\right)}{\Phi\!\left(\tfrac{x}{\sigma_{\rndx}}\right)}, \quad B_{\mathrm{out}}(\gamma)
    \triangleq \Phi\left(\frac{\gamma}{\sigma_{\eff}}\right).
\end{align}
Then,
\tcbset{colframe=gray!90!blue,size=small,width=0.49\textwidth,halign=flush center,arc=2mm,outer arc=1mm}
\begin{tcolorbox}[upperbox=visible,colback=white,halign=left]
\vspace{-0.2cm}
\begin{equation}
\begin{aligned}
    P_{\md}(\gamma)
    &\le \sum_{j=0}^{N-1}
    \int_{0}^{1}
        Q\left(
            \frac{\beta\,\sigma_{\rndx}\Phi^{-1}(u)-\gamma}{\sigma_{\mmse}}
        \right) \\
        &\qquad\qquad\times
        \big[B_{\mathrm{in}}\big(\sigma_{\rndx}\Phi^{-1}(u),\gamma\big)\big]^{M_{\mathrm{in}}(j)}\\
        &\qquad\qquad\times
        \big[B_{\mathrm{out}}(\gamma)\big]^{M_{\mathrm{out}}(j)}
        u^{N-1}\,{\rm d}u. \label{eq:Pmd-bound-u-ct}
\end{aligned}
\end{equation}
\end{tcolorbox}
\end{proposition}

Proposition~\ref{prop2} provides an explicit integral upper bound on $P_{\md}(\gamma)$. However, and while the bound \eqref{eq:Pmd-bound-u-ct} can be (easily) evaluated numerically, for insight into its asymptotic behavior as $k\to\infty$ (so that $N=2^k\to\infty$), it is convenient to examine a simplified approximation thereof, in the spirit of Laplace’s method and extreme-value theory, which we develop next.

\vspace{-0.3cm}
\subsection{Asymptotic Approximation of the MD Upper Bound}
Ignoring the edge indices $j\in\{0,\ldots,L-1\}\cup\{N-L,\ldots,N-1\}$, whose contribution is negligible when $k\to\infty$, the dominant part of the sum in \eqref{eq:Pmd-bound-u-ct} corresponds to interior indices $j\in\{L,\ldots,N-1-L\}$, for which
\begin{equation}\label{eq:setsizesinasymptoticregime}
    M_{\mathrm{in}}(j) = 2L, \qquad M_{\mathrm{out}}(j) = 0.
\end{equation}
For these indices, the integrand in \eqref{eq:Pmd-bound-u-ct} is identical, and the sum over $j$ contributes an overall factor of (approximately) $N$. Using the change of variables $u = \Phi\big(x/\sigma_{\rndx}\big)$, we can rewrite the leading part of the bound in the compact form
\begin{equation}
    P_{\md}(\gamma)
    \lesssim \Exop\Big[
        g\big(\rndx[\rndj]\big)
    \Big],
    \label{eq:Pmd-as-ExpgXmax}
\end{equation}
where
\begin{equation}
    g(x) \triangleq
    Q\!\left( \frac{\beta x - \gamma}{\sigma_{\mmse}} \right)
    \big[B_{\mathrm{in}}(x,\gamma)\big]^{2L},
    \label{eq:def-gx}
\end{equation}
and~\eqref{eq:Pmd-as-ExpgXmax} is exact up to edge effects that vanish as $N\to\infty$.

For large $N$, which grows with $k$ in our scheme, it is well-known from classical extreme-value asymptotics that the maximum $\rndx[\rndj]$ is sharply concentrated around its mean $x_N\triangleq\Exop[\rndx[\rndj]]= \sigma_{\rndx}\sqrt{2\log(N)}(1+o(1))$, and formally,
\begin{equation}
    \lim_{N\to\infty} \Exop\left[ \left(\rndx[\rndj] - x_N\right)^2 \right] = 0.
    \label{eq:xN-asymptotic}
\end{equation}
Thus, more generally, for any \delra{“moderately smooth”}\addra{sufficiently regular} function $g(\cdot)$ one \delra{has}\addra{expects}\addra{\footnote{\addra{We emphasize that \eqref{eq:Laplace-heuristic} is used here as an asymptotic approximation at the logarithmic level, rather than as a sharp relative asymptotic equality. In general, concentration of $\rndx[\rndj]$ around $x_N$ implies that $\Exop[g(\rndx[\rndj])]$ is governed by the behavior of $g(x)$ near $x_N$, but may still differ from $g(x_N)$ by a constant factor. This does not affect the exponent characterized below.}}}
\begin{equation}
    \Exop\big[g(\rndx[\rndj])\big]
    \approx g\left(x_N\right),
    \qquad N\to\infty,
    \label{eq:Laplace-heuristic}
\end{equation}
i.e., the expectation is \delra{dominated}\addra{governed, at the logarithmic scale,} by the \delra{contribution}\addra{behavior} of $g(x)$ near the typical maximum value $x_N$.

Now, for fixed $\gamma, \beta, \sigma_{\rndx}$ and $\sigma_{\eff}$, we have
\begin{align}
    \lim_{x\to\infty}
    \Phi_2\!\left(\tfrac{x}{\sigma_{\rndx}},\tfrac{\gamma}{\sigma_{\eff}};\rho\right)
    &= \Phi\!\left(\tfrac{\gamma}{\sigma_{\eff}}\right)
    = B_{\mathrm{out}}(\gamma), \\
    \lim_{x\to\infty}
    \Phi\left(\tfrac{x}{\sigma_{\rndx}}\right) &= 1
\end{align}
Recalling the definitions \eqref{eq:defofwhat}, it follows that
\begin{equation}
    \lim_{x\to\infty} B_{\mathrm{in}}(x,\gamma) = \Phi\!\left(\tfrac{\gamma}{\sigma_{\eff}}\right) = B_{\mathrm{out}}(\gamma).
    \label{eq:Bin-limit}
\end{equation}
Combining \eqref{eq:xN-asymptotic} and \eqref{eq:Bin-limit}, we obtain
\begin{equation}
    B_{\mathrm{in}}(x_N,\gamma)
    = B_{\mathrm{out}}(\gamma) + o(1),
    \quad N\to\infty,
\end{equation}
and therefore
\begin{equation}
    \big[B_{\mathrm{in}}(x_N,\gamma)\big]^{2L}
    = \big[B_{\mathrm{out}}(\gamma)\big]^{2L}\big(1+o(1)\big),
    \quad N\to\infty.
    \label{eq:Bin-2L-asymptotic}
\end{equation}

Plugging \eqref{eq:def-gx} into \eqref{eq:Laplace-heuristic} and using \eqref{eq:Bin-2L-asymptotic}\delra{, we obtain}\addra{ motivates} the ``large $N$" asymptotic approximation
\begin{align}
    P_{\md}(\gamma)
    &\lesssim g(x_N) \\
    &= Q\!\left(
        \frac{\beta x_N - \gamma}{\sigma_{\mmse}}
    \right)
    \big[B_{\mathrm{in}}(x_N,\gamma)\big]^{2L}\\
    &\approx
    Q\!\left(
        \frac{\beta x_N - \gamma}{\sigma_{\mmse}}
    \right)
    \Big[\Phi\!\left(\tfrac{\gamma}{\sigma_{\eff}}\right)\Big]^{2 \left\lfloor \tfrac{\delta_m}{T_{\mathrm{s}}} \right\rfloor}.
    \label{eq:Pmd-Laplace-approx}
\end{align}
Via the approximation $x_N \approx \sigma_{\rndx}\sqrt{2\log (N)}=\sigma_{\rndx}\sqrt{2k\log (2)}$ that follows from \eqref{eq:xN-asymptotic}, we can further write
\begin{equation}
    P_{\md}(\gamma)
    \lesssim
    Q\!\left(
        \frac{\sigma_{\rnds}\sqrt{2k\log (2)} - \gamma}{\sigma_{\mmse}}
    \right)
    \Big[\Phi\!\left(\tfrac{\gamma}{\sigma_{\eff}}\right)\Big]^{2 \left\lfloor \tfrac{\delta_m}{T_{\mathrm{s}}} \right\rfloor}.
    \label{eq:Pmd-Laplace-approx-simplified}
\end{equation}
\addra{We emphasize that \eqref{eq:Pmd-Laplace-approx}--\eqref{eq:Pmd-Laplace-approx-simplified} are asymptotic approximations intended to provide intuition and scaling insight; the exact nonasymptotic MD upper bound remains \eqref{eq:Pmd-bound-u-ct}, given in Proposition~\ref{prop2}. At the same time,} \delra{T}\addra{t}his asymptotic approximation has three attractive properties:
\begin{itemize}
    \item It is \emph{explicit}: there are no integrals or sums;
    \item It preserves the dependence on $k$ through the Gaussian tail of the aligned term; and
    \item It retains the effect of the local search window through the factor $\big[\Phi(\gamma/\sigma_{\eff})\big]^{2 \left\lfloor \delta_m/ T_{\mathrm{s}} \right\rfloor }$.
\end{itemize}
In particular, \eqref{eq:Pmd-Laplace-approx-simplified} gives the MD error exponent
\begin{equation}
    -\log P_{\md}(\gamma)
    = \frac{\left(\sigma_{\rnds}\sqrt{2k\log(2)} - \gamma\right)^2}{2\sigma_{\mmse}^2}
      + o(k),
    \quad k\to\infty.
\end{equation}

\addra{The asymptotic approximation \eqref{eq:Pmd-Laplace-approx-simplified} also makes the parameter dependence of the MD behavior more interpretable. Specifically, detection improves when the effective signal component becomes stronger relative to the noise, whereas increasing the threshold $\gamma$ makes the detector more conservative and therefore increases the MD probability. Thus, as expected, $\gamma$ controls the standard FA--MD tradeoff: increasing $\gamma$ reduces FA but worsens MD, thereby moving the operating point along the ROC. In addition, the effect of the local search window appears explicitly in the exponent of $\big[\Phi(\gamma/\sigma_{\eff})\big]^{2 \left\lfloor \delta_m/T_{\mathrm{s}} \right\rfloor}$.}

Furthermore, it is interesting to note that, if $\sigma_{\rnds}$ is somehow known to the decoder, it is possible to choose the threshold as a function of $k$ such that the total error probability $\prob{\widehat{\setH}\neq\setH}$ would vanish as $k$ increases, a property that could be thought of as ``communication consistency". Specifically, informed by \eqref{eq:Pmd-Laplace-approx-simplified}, if we let $\gamma(k) = \sigma_{\rnds}h(k)$, such that $h:\naturals\to\positivereals$ and
\begin{align}
    \lim_{k\to\infty}h(k) &= \infty,\\
    \lim_{k\to\infty}\frac{h(k)}{\sqrt{k}} &= 0,
\end{align}
denoting $\mybar{P}_{\fa}(\gamma)$ and $\mybar{P}_{\md}(\gamma)$ as the upper bounds \eqref{eq:falsealaramupperbound} and \eqref{eq:Pmd-bound-u-ct}, respectively, we readily have
\begin{equation}
    \lim_{k\to\infty} \prob{\widehat{\setH}\neq\setH} \leq \lim_{k\to\infty} \left[ \mybar{P}_{\fa}\left(\gamma(k)\right) + \mybar{P}_{\md}\left(\gamma(k)\right) \right] = 0.
\end{equation}
Thus, with the additional parametric knowledge of $\sigma_{\rnds}$, our proposed detector can be made arbitrarily accurate with the communication budget.

\addra{ The subsections in the remainder of this section are designated exclusively for the proofs of Propositions~\ref{prop1} and~\ref{prop2}.}

\subsection{Proof of Proposition \ref{prop1}}\label{subsec:falsealaramprob}
The FA event corresponds to the process $\rndy(t)$, which under $\setH_0$ is equal to $\rndz_2(t)$, exceeding a fixed threshold $\gamma > 0$ at any time in the delay window $\Delta = [-\delta_m, \delta_m]$ . Therefore,
\begin{equation}
    P_{\fa} = \prob{ \sup_{\tau \in \Delta} \rndy(t_{\rndj} + \tau) > \gamma \mid \setH_0 } = \prob{ \sup_{t \in \Delta} \rndz_2(t) > \gamma },
\end{equation}
and note that $ \left\{ \sup_{t \in \Delta} \rndz_2(t) > \gamma \right\} $ can be partitioned such that
\begin{align}
P_{\fa} &= \prob{\rndz_2(-\delta_m) > \gamma} \label{eq:twoevent1}\\
&+ \prob{ \sup_{t \in (-\delta_m, \delta_m]} \rndz_2(t) > \gamma,\ \rndz_2(-\delta_m) \leq \gamma }. \label{eq:twoevent2}
\end{align}

Define the RV $ \rndn_{\gamma}\in\naturals $ as the number of upcrossings of level \( \gamma \) by \( \rndz_2(t) \) over \( [-\delta_m, \delta_m] \). Thus, using Markov's inequality, \eqref{eq:twoevent2} is upper bounded by
\begin{align}
    \hspace{-0.1cm}\prob{ \sup_{t \in (-\delta_m, \delta_m]} \rndz_2(t) > \gamma,\ \rndz_2(-\delta_m) \leq \gamma }
    &\leq \mathbb{P}(\rndn_{\gamma} \geq 1) \\
    &\leq \Exop[\rndn_{\gamma}].
\end{align}
It follows that the FA probability is bounded from above as
\begin{equation}\label{eq:simpleformoffaupperbound}
    P_{\fa} \leq \prob{\rndz_2(-\delta_m) > \gamma} + \Exop[\rndn_{\gamma}].
\end{equation}

For stationary Gaussian processes with differentiable sample paths, we have from Rice's formula (e.g.,~\cite{ito1963expected})
\begin{equation}\label{eq:riceformula}
    \mathbb{E}[\rndn_{\gamma}] = \frac{\delta_m}{\pi} \cdot \frac{\sqrt{-R''_{\rndz_2\rndz_2}(0)}}{\sigma_2} \cdot \exp\left( -\frac{\gamma^2}{2\sigma_2^2} \right),
\end{equation}
where $ R''_{\rndz_2\rndz_2}(0) = \left.R''_{\rndz_2\rndz_2}(\tau)\right|_{\tau=0}$ is the second derivative (with respect to $\tau$) of the autocorrelation function $R_{\rndz_2\rndz_2}(\tau)$ evaluated at zero. By the Wiener--Khinchin theorem, computing $ R''_{\rndz_2\rndz_2}(0) $ explicitly using the PSD \eqref{eq:psdnoise}, we have
\begin{align}
    R_{\rndz_2\rndz_2}''(0) &= -4\pi^2 \int_{-\infty}^{\infty} f^2 S_{\rndz_2\rndz_2}(f) {\rm d}f \\
    &= -4\pi^2\cdot\frac{\sigma_2^2}{B} \cdot \int_{-B/2}^{B/2} f^2 {\rm d}f =-\frac{\pi^2\sigma_2^2 B^2}{3}.
\end{align}
Substituting this into \eqref{eq:riceformula} and further simplifying gives
\begin{align}\label{eq:ricesimplified}
    \mathbb{E}[\rndn_{\gamma}] &= \frac{\delta_m B}{\sqrt{3}} \cdot \exp\left( -\frac{\gamma^2}{2\sigma_2^2} \right).
\end{align}
Now, substituting $\mathbb{P}(\rndz_2(-\delta_m) > \gamma)=Q\left( \frac{\gamma}{\sigma_2} \right)$ and \eqref{eq:ricesimplified} into \eqref{eq:simpleformoffaupperbound}, gives \eqref{eq:falsealaramupperbound}, which completes the proof of the proposition.

\subsection{Proof of Proposition \ref{prop2}}\label{subsec:misdetectionprob}
Recall that under $\setH_1$, the signals at the encoder and decoder are given by \eqref{eq:h1}, which in particular implies
\begin{align}
    \rndx(t-\delta) &= \rnds(t-\delta) + \rndz_1(t-\delta)\\
    \Rightarrow \; \rndy(t) &= \rndx(t-\delta) + \underbrace{\rndz_2(t) - \rndz_1(t-\delta)}_{\triangleq\rndw(t)}, \label{eq:representationofy1}
\end{align}
and, by definition, $\rndw(t)$ is Gaussian and correlated with $\rndx(t)$. A well-known property of two correlated jointly Gaussian RVs is that one can be expressed as a scaled version of the other plus an \emph{independent} Gaussian RV, which are precisely the minimum mean square error (MMSE) estimator and the corresponding residual estimation error, respectively. Leveraging this property, $\rndw(t)$ can be represented as
\begin{equation}\label{eq:wtmmse}
    \rndw(t) = (\beta-1)\rndx(t-\delta) + \mybar{\rndz}(t),
\end{equation}
where $\mybar{\rndz}(t)\sim\normal(0,\sigma^2_{\mmse})$ is a stationary process with PSD $S_{\bar{\rndz}\bar{\rndz}}(f) = \frac{\sigma^2_{\mmse}}{B} \rectf\left(\frac{f}{B}\right)$, independent of $\rndx(t)$, and $\beta$ and $\sigma^2_{\mmse}$ are defined in \eqref{eq:defofmmseconstants}--\eqref{eq:defofmmseconstants2}. Having established the representation \eqref{eq:wtmmse}, substituting it into \eqref{eq:representationofy1} yields
\begin{align}
    \rndy(t) = \beta\rndx(t-\delta) + \mybar{\rndz}(t). \label{eq:representationofy2}
\end{align}

Now, define $\rndy_{\delta}[n]\triangleq \rndy(nT_{\mathrm{s}}+\delta)$ for all $n\in\integers$, the set 
\begin{equation}\label{eq:countableset}
    \Delta_{t_{\rndj}}\triangleq \{k\in\integers\,\mid\,(kT_{\mathrm{s}})\in [t_{\rndj}-\delta_m,t_{\rndj}+\delta_m]\},
\end{equation}
and
\begin{equation}
     \rndy_{\delta}^{\max} \triangleq \max_{n\in \Delta_{t_{\rndj}}} \; \rndy_\delta[n],
    \label{eq:maxofyforproof}
\end{equation}
Then, since $\rndy_{\delta}^{\max} \leq \sup_{\tau \in \Delta} \rndy(t_{\rndj} + \tau)$, it holds that
\begin{equation}
P_{\md}\leq\prob{ \rndy_{\delta}^{\max} < \gamma \mid \setH_1 }.\label{eq:firstupperboundmd}
\end{equation}
Further developing the upper bound \eqref{eq:firstupperboundmd}, we have
\begin{align}\label{eq:secondupperboundmd}
    &\prob{ \rndy_{\delta}^{\max} < \gamma \mid \setH_1 }\\
    &=\Exop\left[\prob{ \rndy_{\delta}^{\max} < \gamma \mid \setX_N,\setH_1 }\mid \setH_1\right] \label{eq:firsttimelote}\\
    &=\Exop\left[\prod_{n\in \Delta_{t_{\rndj}}}\prob{  \mybar{\rndz}_{\delta}[n]< \gamma-\beta\rndx[n] \mid \setX_N,\setH_1 }\mid \setH_1\right] \label{eq:normalnoiseresidualsindependence}\\
    &=\Exop\left[\prod_{n\in \Delta_{t_{\rndj}}}Q\left( \frac{\beta\rndx[n]-\gamma}{\sigma_{\mmse}} \right)\mid \setH_1\right]\label{eq:normalnoiseresiduals}\\
    &=\Exop\left[Q\left( \frac{\beta\rndx[\rndj]-\gamma}{\sigma_{\mmse}} \right)\left.Q^{D-1}\left( \frac{\beta\rndx[n]-\gamma}{\sigma_{\mmse}} \right)\right|_{n\neq\rndj}\mid \setH_1\right]\label{eq:truncatedareaindependent},
\end{align}
where  $D\triangleq \left|\Delta_{t_{\rndj}}\right|\in\naturals$, and we have used:
\begin{itemize}
    \item the law of total expectation in \eqref{eq:firsttimelote};
    \item that $\{\mybar{\rndz}_{\delta}[n]\triangleq\mybar{\rndz}(nT_{\mathrm{s}}+\delta)\sim\normal(0,\sigma^2_{\mmse})\}_{n\in\integers}$ are independent in \eqref{eq:normalnoiseresidualsindependence} and identically distributed  in \eqref{eq:normalnoiseresiduals}; and
    \item $\{\rndx[n]\}_{n\neq\rndj}$, and hence $\{Q\left( \tfrac{\beta\rndx[n]-\gamma}{\sigma_{\mmse}} \right)\}_{n\neq\rndj}$ are independent, identically distributed (i.i.d.) given $(\rndj,\rndx[\rndj], \setH_1)$ in \eqref{eq:truncatedareaindependent}.
\end{itemize}
We note that while the set $\Delta_{t_{\rndj}}$ is random, its cardinality $D$ is deterministic (for details, see~\eqref{eq:appcardinalityD1}--\eqref{eq:appcardinalityD2}, Appendix~\ref{app:explicittermscardinality}).

Continuing from \eqref{eq:truncatedareaindependent}, we further develop the upper bound by conditioning on the random pair $(\rndj,\rndx[\rndj])$ and exploiting the structure of the discrete-time samples. Recall that, due to the rectangular PSDs and Nyquist sampling, $\{\rndx[n]\}_{n\in\integers}$ are i.i.d.\ Gaussian with variance $\sigma_{\rndx}^2$, and under $\setH_1$ it holds that
\begin{equation}
    \rndy_{\delta}[n] \triangleq \rndy(nT_{\mathrm{s}}+\delta) = \beta\rndx[n] + \mybar{\rndz}_{\delta}[n],
\end{equation}
where $\mybar{\rndz}_{\delta}[n]\overset{\text{iid}}{\sim}\normal(0,\sigma_{\mmse}^2)$ is independent of $\{\rndx[n]\}_{n\in\integers}$.

By symmetry, $\rndj$ is uniform on $\{0,\ldots,N-1\}$ and independent of $\rndx[\rndj]$. Moreover, conditioned on $\{\rndj=j,\rndx[\rndj]=x\}$, the $N-1$ remaining samples are i.i.d.\ as $\rndv\mid \rndv\le x$ with $\rndv\sim\normal(0,\sigma_{\rndx}^2)$. Therefore, we may rewrite \eqref{eq:truncatedareaindependent} as
\begin{align}
    &\prob{ \rndy_{\delta}^{\max} < \gamma \mid \setH_1 } \\
    &= \Exop\left[
        Q\left( \tfrac{\beta\rndx[\rndj]-\gamma}{\sigma_{\mmse}} \right)
        \Exop\left[
            \prod_{\substack{n\in\Delta_{t_{\rndj}} \\ n\neq\rndj}}
            Q\left( \tfrac{\beta\rndx[n]-\gamma}{\sigma_{\mmse}} \right)
            \Bigm| \rndj,\rndx[\rndj],\setH_1
        \right]
        \Bigm| \setH_1
    \right]. \label{eq:cond-on-j-and-max}
\end{align}

Next, for a fixed realization $\{\rndj=j,\rndx[\rndj]=x\}$, we decompose the non-aligned indices $n\in\Delta_{t_j}\setminus\{j\}$ into ``inside'' and ``outside'' lags relative to the encoded block $\{0,\ldots,N-1\}$:
\begin{align}
    \Delta_{t_j}^{\mathrm{in}} &\triangleq \big(\Delta_{t_j}\cap\{0,\ldots,N-1\}\big)\setminus\{j\}, \label{eq:inside-set}\\
    \Delta_{t_j}^{\mathrm{out}} &\triangleq \Delta_{t_j}\setminus\big(\Delta_{t_j}^{\mathrm{in}}\cup\{j\}\big), \label{eq:outside-set}
\end{align}
and define their cardinalities
\begin{equation}
    M_{\mathrm{in}}(j) \triangleq \big|\Delta_{t_j}^{\mathrm{in}}\big|,
    \quad
    M_{\mathrm{out}}(j) \triangleq \big|\Delta_{t_j}^{\mathrm{out}}\big|
    = D-1-M_{\mathrm{in}}(j), \label{eq:numberofinsideoutside}
\end{equation}
where we recall that $D=|\Delta_{t_{\rndj}}|$.

\paragraph*{Outside lags}
For $n\in\Delta_{t_j}^{\mathrm{out}}$, the corresponding samples $\rndx[n]$ lie outside the encoded block and are therefore independent of $\setX_N$ (and of $(\rndj,\rndx[\rndj])$), with law $\normal(0,\sigma_{\rndx}^2)$. Let $\rndv\sim\normal(0,\sigma_{\rndx}^2)$ and $\rndg\sim\normal(0,1)$ be independent, and define
\begin{equation}
    \rndu \triangleq \beta\,\rndv - \sigma_{\mmse}\rndg
    \sim\normal\left(0,\sigma_{\eff}^2\right). \label{eq:sigmaeff-def}
\end{equation}
Then, for a generic outside lag,
\begin{align}
    B_{\mathrm{out}}(\gamma)
    &\triangleq
    \Exop\left[
        Q\left(\frac{\beta\rndx[n]-\gamma}{\sigma_{\mmse}}\right)
        \Bigm| \rndj=j,\rndx[\rndj]=x,\setH_1
    \right] \\
    &= \Exop\left[
        Q\left(\frac{\beta\rndv-\gamma}{\sigma_{\mmse}}\right)
    \right] \\
    &= \Exop\left[
        \prob{\rndg > \tfrac{\beta\rndv-\gamma}{\sigma_{\mmse}} \mid \rndv}
    \right]
     = \prob{\beta\rndv - \sigma_{\mmse}\rndg < \gamma} \\
    &= \prob{\rndu<\gamma}
     = \Phi\left(\frac{\gamma}{\sigma_{\eff}}\right). \label{eq:Bout-ct}
\end{align}

\paragraph*{Inside lags}
For $n\in\Delta_{t_j}^{\mathrm{in}}$, the samples $\rndx[n]$ lie within the encoded block and are affected by the conditioning on the maximum. Specifically, conditioned on $\{\rndj=j,\rndx[\rndj]=x\}$, the $M_{\mathrm{in}}(j)$ ``inside samples" are i.i.d.\ as $\rndv\mid\rndv\le x$, where $\rndv\sim\normal(0,\sigma_{\rndx}^2)$. Therefore, for a generic inside lag,
\begin{align}
    B_{\mathrm{in}}(x,\gamma)
    &\triangleq
    \Exop\left[
        Q\left(\frac{\beta\rndx[n]-\gamma}{\sigma_{\mmse}}\right)
        \Bigm| \rndj=j,\rndx[\rndj]=x,\setH_1
    \right] \\
    &= \Exop\left[
        Q\left(\frac{\beta\rndv-\gamma}{\sigma_{\mmse}}\right)
        \Bigm| \rndv\le x
    \right] \\
    &= \prob{\rndu<\gamma \mid \rndv\le x}. \label{eq:Bin-def-ct}
\end{align}
Let $\rho$ denote the correlation coefficient between $\rndv$ and $\rndu$,
\begin{equation}
    \rho \triangleq \frac{\Exop[\rndv\rndu]}{\sigma_{\rndx}\sigma_{\eff}}
    = \frac{\beta\sigma_{\rndx}}{\sigma_{\eff}}. \label{eq:rho-def-ct}
\end{equation}
Then, by Bayes’ rule,
\begin{align}
    B_{\mathrm{in}}(x,\gamma)
    &= \frac{\prob{\rndu<\gamma,\;\rndv\le x}}{\prob{\rndv\le x}} = \frac{
        \Phi_2\!\left(\tfrac{x}{\sigma_{\rndx}},\tfrac{\gamma}{\sigma_{\eff}};\rho\right)
    }{
        \Phi\!\left(\tfrac{x}{\sigma_{\rndx}}\right)
    }. \label{eq:Bin-ct}
\end{align}

Since, conditioned on $\{\rndj=j,\rndx[\rndj]=x\}$, the inside and outside samples are independent, the product in \eqref{eq:cond-on-j-and-max} reads
\begin{align}
    &\Exop\left[
        \prod_{\substack{n\in\Delta_{t_{j}} \\ n\neq j}}
        Q\left( \frac{\beta\rndx[n]-\gamma}{\sigma_{\mmse}} \right)
        \Bigm| \rndj=j,\rndx[\rndj]=x,\setH_1
    \right] \\
    &\qquad=
    \left[B_{\mathrm{in}}(x,\gamma)\right]^{M_{\mathrm{in}}(j)}
    \left[B_{\mathrm{out}}(\gamma)\right]^{M_{\mathrm{out}}(j)}. \label{eq:insideoutside-product-ct}
\end{align}

Substituting \eqref{eq:insideoutside-product-ct} into \eqref{eq:cond-on-j-and-max}, we obtain
\begin{align}
    &\prob{ \rndy_{\delta}^{\max} < \gamma \mid \setH_1 }\\
    &= \Exop\Bigg[
        Q\left( \tfrac{\beta\rndx[\rndj]-\gamma}{\sigma_{\mmse}} \right)
        \left[B_{\mathrm{in}}\big(\rndx[\rndj],\gamma\big)\right]^{M_{\mathrm{in}}(\rndj)}
        \left[B_{\mathrm{out}}(\gamma)\right]^{M_{\mathrm{out}}(\rndj)}
        \Bigm| \setH_1
    \Bigg]. \label{eq:Pmd-bound-expectation-ct}
\end{align}
As mentioned above, $\rndj$ is uniformly distributed on $\{0,\ldots,N-1\}$, and the probability density function of the block maximum $\rndx[\rndj]$ is given by the classical order-statistics formula
\begin{equation}
    p_{\rndx[\rndj]}(x)
    = \frac{N}{\sigma_{\rndx}}\,
    \phi\!\left(\frac{x}{\sigma_{\rndx}}\right)\,
    \Phi\!\left(\frac{x}{\sigma_{\rndx}}\right)^{N-1},
    \qquad x\in\reals. \label{eq:pdf-max-ct}
\end{equation}
Thus, the upper bound on the MD probability can be written explicitly in closed-form as
\begin{align}
    P_{\md}(\gamma)
    &\le \prob{ \rndy_{\delta}^{\max} < \gamma \mid \setH_1 } \\
    &= \frac{1}{N}\sum_{j=0}^{N-1}
    \int_{-\infty}^{\infty}
        \underbrace{
        Q\left(\frac{\beta\,x-\gamma}{\sigma_{\mmse}}\right)
        }_{\text{aligned}}
        \cdot
        \underbrace{
        \big[B_{\mathrm{in}}(x,\gamma)\big]^{M_{\mathrm{in}}(j)}
        }_{\text{inside lags}}
        \\
        &\qquad\qquad\quad\;\;\times\underbrace{
        \big[B_{\mathrm{out}}(\gamma)\big]^{M_{\mathrm{out}}(j)}
        }_{\text{outside lags}} \,\cdot\, p_{\rndx[\rndj]}(x)\,{\rm d}x. \label{eq:Pmd-bound-integral-ct}
\end{align}

Using the change of variables $u=\Phi\!\left(x/\sigma_{\rndx}\right)$, hence $x=\sigma_{\rndx}\Phi^{-1}(u)$ and $p_{\rndx[\rndj]}(x)\,{\rm d}x = N\,u^{N-1}{\rm d}u$, \eqref{eq:Pmd-bound-integral-ct} can be expressed as \eqref{eq:Pmd-bound-u-ct}. To complete the characterization of the bound, in Appendix~\ref{app:explicittermscardinality} the dependence of $D$, $M_{\mathrm{in}}(j)$, and $M_{\mathrm{out}}(j)$ on the model parameters is made explicit via \eqref{eq:minhowmuch}--\eqref{eq:mouthowmuch}.

\section{Simulation Results}\label{sec:simulation}
In this section we validate our analytical results for our scheme, and specifically our \textbf{MID} detector \eqref{eq:detector_rule_ct}. While doing so, we also compare the performance of our scheme to:
\begin{itemize}
    \item a realizable benchmark based on the commonly used one-bit-per-sample (scalar) quantization, which has been gaining renewed interest for various related tasks (e.g.,~\cite{huang2019one,bhandari2020one,weiss2021one,eamaz2024uno,kumar2025carrier}), denoted as \textbf{1-bit}. In this scheme,
    \begin{align}
        \rvecm{\onebit} &\triangleq \tp{[\sign(\rndx[0])\, \cdots \, \sign(\rndx[k-1])]},\\
        \rndt_{\onebit} &\triangleq \max_{\tau \in \Delta} \; \widehat{R}_{\onebit}(\tau),
    \end{align}
    are the $k$-bit message\addra{, i.e., the signs of the first $k$ consecutive Nyquist-rate samples observed at the encoder,} and the constructed statistic, respectively, where $\rndt_{\onebit}$ is compared to a threshold and
    \begin{align}
    \widehat{R}_{\onebit}(\tau)  &\triangleq \displaystyle\int_0^T \widetilde{\rndx}_{\onebit}(t)\rndy(t+\tau)\,\mathrm{d}t,\label{eq:corr_unnormalized}\\
    \widetilde{\rndx}_{\onebit}(t)
    &\triangleq
    \sum_{n=0}^{k-1}
        \sign\left(\rndx[n]\right)\cdot
        \mathbbm{1}_{t\in[nT_{\mathrm{s}},\,(n+1)T_{\mathrm{s}})}.
\end{align}
\addra{\item a transform-domain benchmark (realizable as well) inspired by Fowler and Chen's Fisher-information (FI)-based compression for TDE~\cite{fowler2005fisher}, denoted as \textbf{FI}. Since the original FI-based formulation is estimation-oriented, we use here a detection-oriented adaptation. In this scheme,
\begin{align}
    \rvecm_{\fisherinf} &\triangleq \tp{\left[\rvecm_{\fisherinf,{\indsub}}\;\rvecm_{\fisherinf,{\quant}}\right]},
\end{align}
where $\rvecm_{\fisherinf,{\indsub}}\in\{0,1\}^{k_{\indsub}\times1}$ encodes the indices of the retained DFT coefficients selected according to their FI for TDE (such that $k_{\indsub}<k$), and $\rvecm_{\fisherinf,{\quant}}\in\{0,1\}^{(k-k_{\indsub})\times1}$ encodes their associated quantized values.\footnote{\addra{Since the regime considered here is extremely rate-limited ($k\sim 10$ bits), we use a simplified ultra-low-rate implementation. Specifically, for even $k$, we set the DFT length to $N=2^{k/2}$ and split the total $k$-bit budget equally: $k/2$ bits are used to identify the retained DFT coefficient and the remaining $k/2$ bits are used to quantize its value (when $k$ is odd, the ``extra" bit goes to quantization). This benchmark is inspired by the FI-based transform-domain scheme for TDE proposed by Fowler and Chen~\cite{fowler2005fisher}, although it is not identical, as their (estimation, not detection) scheme does not appear to scale favorably down to the extremely low-bit-budget regime considered here.}} The corresponding compressed waveform $\widetilde{\rndx}_{\fisherinf}(t)$ is reconstructed from these retained coefficients, and the resulting delay-search score
\begin{align}
    \widehat{R}_{\fisherinf}(\tau)
    &\triangleq
    \int_0^T \widetilde{\rndx}_{\fisherinf}(t)\rndy(t+\tau)\,\mathrm{d}t,
\end{align}
is used to construct the statistic
\begin{align}
    \rndt_{\fisherinf} &\triangleq \max_{\tau \in \Delta} \; \widehat{R}_{\fisherinf}(\tau),
\end{align}
which is then compared to a threshold.}
    \item a theoretical benchmark based on the optimal rate-distortion compression for the signal $\rndx(t)$, where the distortion measure is the squared error $\left(\rndx(t)-\widehat{\rndx}_{\ratedist}(t)\right)^2$, denoted as \textbf{RD}. In this case, the test \eqref{eq:ct_glrt_corr} with $\rndx(t)$ replaced by $\widehat{\rndx}_{\ratedist}(t)$, after appropriately setting the threshold, is the still the GLRT since $\widehat{\rndx}_{\ratedist}(t)$ is Gaussian and jointly Gaussian with $\rndy(t)$ (see, e.g.,~\cite{kolmogorov1956shannon,ihara1993information}). Nevertheless, in our simulations, for RD we use \eqref{eq:corr_unnormalized} with $\widetilde{\rndx}_{\onebit}(t)$ replaced by $\widehat{\rndx}_{\ratedist}(t)$ since it exhibited better performance.\footnote{Indeed, the GLRT is theoretically justified only in the asymptotic regime, and is generally \emph{not} optimal nonasymptotically. For this reason, we use the unnormalized statistic \eqref{eq:corr_unnormalized} for 1-bit as well, which performed better.}
\end{itemize}

Throughout, the signals $\rnds(t),\rndz_1(t)$ and $\rndz_2(t)$ are generated according to \eqref{eq:h0}--\eqref{eq:h1} with $B = 1$ Hz, where $\sigma_{\rnds}=1$ and $T_{\mathrm{s}} = 1/B$ sec. The observation interval is of length $T = N T_{\mathrm{s}}$, where $N=2^k$. To simulate the continuous-time bandlimited model, we use an oversampled discrete-time approximation. Specifically, we first generate i.i.d.\ Gaussian sequences, corresponding to the discrete-time samples of $\rnds(t)$ and the noise processes sampled at the Nyquist rate (sampling period $T_{\mathrm{s}}$). We then upsample each sequence by a factor $R_{\mathrm{up}}$ via zero insertion and apply a lowpass, truncated-sinc interpolation filter to obtain a fine-grid representation of the underlying bandlimited waveforms with sampling period $T_{\mathrm{s}}/R_{\mathrm{up}}$. Delays are then applied on this fine grid. The encoder, however, only observes Nyquist-rate samples (every $R_{\mathrm{up}}$-th fine-grid sample) over a block of length $N$, whereas the decoder evaluates the relevant decision statistic by scanning over all the fine-grid delays within $\Delta$. All empirical results (1-bit, RD, MID) were obtained based on $10^5$ independent trials.

\subsection{ROC Curve Evaluation}
\begin{figure}[t]
    \centering
    
        \begin{tikzpicture}
            \begin{axis}[
                width=\linewidth,
                height=0.8*\linewidth,
                axis on top = false,
                grid=major,
                xlabel={$P_{\fa}$},
                ylabel={$P_{\detc}$},
                xmin=0, xmax=1,
                ymin=0, ymax=1,
                xtick={0,0.2,0.4,0.6,0.8,1},
                ytick={0,0.2,0.4,0.6,0.8,1},
                legend pos=south east,
                legend cell align={left},
                legend style={font=\footnotesize},
                legend style={
                    at={(0.99,0.01)},
                    anchor=south east},
                axis on top,
                domain=0:1,
            ]

            \addplot [
                dashdotdotted,
                color=blue,
                mark size=1.5pt,
                line width=1.5pt,
                opacity=0.8,
            ]
            table [x index=0, y index=1] {data/ROC_1b.dat};

            \addplot [
                loosely dotted,
                color=magenta,
                mark size=1.5pt,
                line width=1.5pt,
                opacity=0.8,
            ]
            table [x index=0, y index=1] {data/fig2_ROC_fowler_fair.dat};

            \addplot [
                dashed,
                color=purple,
                mark size=1.5pt,
                line width=1.5pt,
                opacity=0.7,
            ]
            table [x index=0, y index=1] {data/ROC_RD.dat};

            \addplot [
                color=teal,
                mark size=1.5pt,
                line width=1.5pt,
                opacity=0.8,
            ]
            table [x index=0, y index=1] {data/ROC_ext.dat};

            \addplot [
                dashdotted,
                color=uppboundcolor,
                mark size=1.5pt,
                line width=1.5pt,
                opacity=0.8,
            ]
            table [x index=0, y index=1] {data/ROC_bound.dat};

            \addplot [
                dotted,
                color=brown,
                line width=1.5pt,
                opacity=0.8,
            ]
            table [x index=0, y index=1] {data/ROC_bound_lap.dat};

            \legend{1-bit, FI, RD, MID, Exact Bound~\eqref{eq:Pmd-bound-u-ct}, Approximated Bound~\eqref{eq:Pmd-Laplace-approx-simplified}}
            \end{axis}
        \end{tikzpicture}\vspace{-0.2cm}

    \caption{ROC curves of 1-bit, RD and MID for $k=8$~bits, $N=256$~Nyquist samples, $\delta_m=200$~sec and $\mathsf{SNR}_{\rndx}=\mathsf{SNR}_{\rndy}=0$~dB. MID is superior to the practical benchmark 1-bit, and slightly outperforms the theoretical benchmark RD. The bounds \eqref{eq:falsealaramupperbound} and \eqref{eq:Pmd-bound-u-ct} are seen to be valid, albeit loose for some FA probabilities; the approximated bound \eqref{eq:Pmd-Laplace-approx-simplified} is seen to be reliable.}
    \label{fig:roc_combined}\vspace{-0.3cm}
\end{figure}
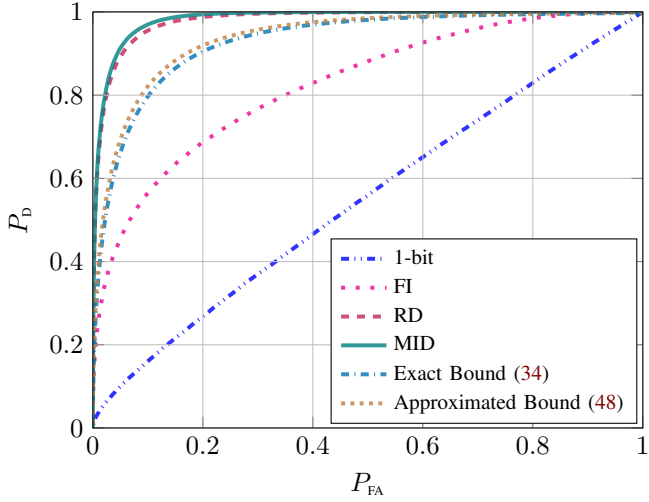
In our first experiment we evaluate the empirical receiver operating characteristic (ROC) of our MID, and compare it with 1-bit and RD (as detailed above) in a representative practical nonasymptotic setting. The ROC curves are computed by sweeping the thresholds and estimating the resulting FA and MD (equivalently, detection $P_{\detc}\triangleq 1-P_{\md}$) probabilities in a Monte-Carlo fashion. For this experiment, we set $k=8$ bits, $N=256$ Nyquist samples, $\delta_m=200$ sec as the maximum admissible delay, and $\mathsf{SNR}_{\rndy}=\mathsf{SNR}_{\rndx}=0$ dB. Notice that in this setting, $\delta_m$ is $\sim\hspace{-0.1cm}78\%$ of $T$, hence $M_{\mathrm{out}}(\rndj)$ is highly unlikely to be zero (unlike in the asymptotic regime~\eqref{eq:setsizesinasymptoticregime}).

Fig.~\ref{fig:roc_combined} shows that the proposed MID detector significantly outperforms 1-bit\addra{ and FI}, and even slightly improves upon (the nonrealizable) RD theoretical benchmark. We emphasize that this improvement over RD is theoretically possible, since optimal RD compression with respect to a \emph{signal reconstruction} distortion measure followed by the GLRT in general does \emph{not} guarantee to be optimal in the sense \eqref{eq:optimizationproblem}.

We also plot a ROC curve based on our upper bound\addra{s}, which can be seen to be valid, but somewhat loose in this nonasymptotic regime.\addra{ We note that this looseness should not come as a surprise: Propositions~\ref{prop1} and~\ref{prop2} provide exact, nonasymptotic upper bounds for events governed by extrema of a continuous-time Gaussian process over a delay window, and sharp finite-sample supremum distributions in such settings are classically difficult except in very special cases (see, e.g.,~\cite{adler2007random,piterbarg1996asymptotic,azais2009level}). In the considered nonasymptotic regime, $k=8$ is still modest and the delay window is relatively large, so the resulting bound remains conservative.} We further plot the bound-based ROC while using the approximated, and simplified, upper bound \eqref{eq:Pmd-Laplace-approx-simplified}, which is seen to be fairly reliable relative to the exact upper bound. These results support the theoretical insight that leveraging the extremum location can yield high detection reliability even under such tight bit constraints. Across different choices of $k, \delta_m$ and SNR levels, the ROC curves exhibit the same qualitative behavior.

\subsection{Error Probability versus SNR and Bit Budget}\label{subsec:pe_vs_k}
In our second experiment we demonstrate how the overall detection reliability improves as the communication budget increases, under a fixed FA constraint. Specifically, we consider
\begin{equation}
     \prob{\widehat{\setH} \neq \setH } = \tfrac{1}{2} P_{\fa} + \tfrac{1}{2} P_{\md},
\end{equation}
and enforce a common design level $P_{\fa}=\epsilon_{\fa}$ for all schemes---for each detector, we empirically determine, under $\mathcal{H}_0$, a decision threshold that yields $\epsilon_{\fa}$. We then fix this threshold and estimate the corresponding $P_{\detc}$ under $\mathcal{H}_1$. This Neyman--Pearson–type construction guarantees a fair comparison, since all curves are evaluated at the same FA level.
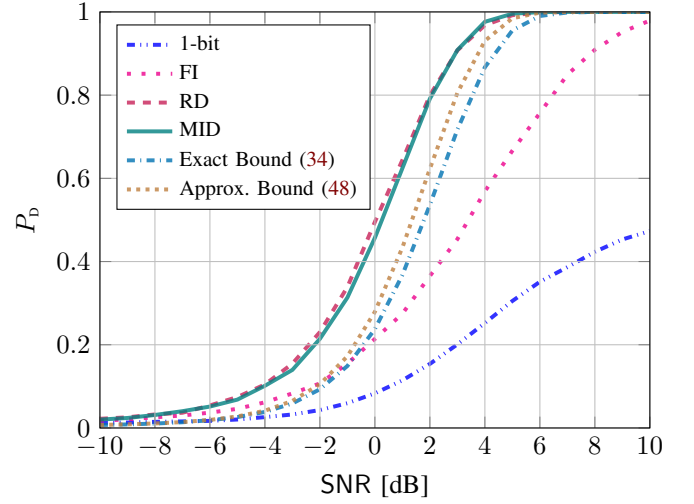
\begin{figure}[t]
    \centering
    
        \begin{tikzpicture}
            \begin{axis}[
                width=\linewidth,
                height=0.8*\linewidth,
                axis on top = false,
                grid=major,
                xlabel={$\mathsf{SNR}$ [dB]},
                ylabel={$P_{\detc}$},
                xmin=-10, xmax=10,
                ymin=0, ymax=1,
                xtick={-10,-8,-6,-4,-2,0,2,4,6,8,10},
                ytick={0,0.2,0.4,0.6,0.8,1},
                legend pos=north west,
                legend cell align={left},
                legend style={font=\footnotesize},
                axis on top,
                domain=0:1,
            ]

            \addplot [
                dashdotdotted,
                color=blue,
                mark size=1.5pt,
                line width=1.5pt,
                opacity=0.8,
            ]
            table [x index=0, y index=1] {data/pd_vs_snr_1b.dat};

            \addplot [
                loosely dotted,
                color=magenta,
                mark size=1.5pt,
                line width=1.5pt,
                opacity=0.8,
            ]
            table [x index=0, y index=1] {data/fig3_pd_vs_snr_fowler_fair.dat};

            \addplot [
                dashed,
                color=purple,
                mark size=1.5pt,
                line width=1.5pt,
                opacity=0.7,
            ]
            table [x index=0, y index=1] {data/pd_vs_snr_RD.dat};

            \addplot [
                color=teal,
                mark size=1.5pt,
                line width=1.5pt,
                opacity=0.8,
            ]
            table [x index=0, y index=1] {data/pd_vs_snr_ext.dat};

            \addplot [
                dashdotted,
                color=uppboundcolor,
                mark size=1.5pt,
                line width=1.5pt,
                opacity=0.8,
            ]
            table [x index=0, y index=1] {data/pd_vs_snr_bound_exact.dat};

            \addplot [
                dotted,
                color=brown,
                line width=1.5pt,
                opacity=0.8,
            ]
            table [x index=0, y index=1] {data/pd_vs_snr_bound_approx.dat};

            \legend{1-bit, FI, RD, MID, Exact Bound~\eqref{eq:Pmd-bound-u-ct}, Approx\addra{.}\delra{imated} Bound~\eqref{eq:Pmd-Laplace-approx-simplified}}
            \end{axis}
        \end{tikzpicture}\vspace{-0.2cm}

    \caption{Probability of detection versus SNR, where $\mathsf{SNR}=\mathsf{SNR}_{\rndx}=\mathsf{SNR}_{\rndy}$, for $k=7$~bits, $N=128$~Nyquist samples, $\delta_m=60$~sec, and a fixed $\epsilon_{\fa}=10^{-2}$. \delra{The legend in Fig.~\ref{fig:roc_combined} applies to Fig.~\ref{fig:pd_vs_snr} as well. }Here, MID nearly attains the theoretical RD performance, while considerably outperforming the 1-bit practical benchmark. The exact and approximated bounds provide a conservative prediction, but with an accurate trend.}
    \label{fig:pd_vs_snr}\vspace{-0.4cm}
\end{figure}

We first fix $k=7$ bits, the maximum admissible delay $\delta_m=60$~sec, and plot the resulting probability of detection, i.e., $P_{\rm D}$ at fixed $\epsilon_{\fa}=10^{-2}$, as a function of the common SNR level, namely $\mathsf{SNR}=\mathsf{SNR}_{\rndx}=\mathsf{SNR}_{\rndy}$. Fig.~\ref{fig:pd_vs_snr} shows that MID exhibits a clear and consistent improvement as $\mathsf{SNR}$ grows, in accordance with the expectation of this natural behavior. However, the 1-bit scheme provides only limited gains with increasing $\mathsf{SNR}$ in this regime, reflecting the fact that scalar quantization is far from optimal, and perhaps a na\"ive compression strategy in this detection-oriented setting.\addra{ The FI benchmark, while substantially better than 1-bit, also remains considerably inferior to MID, indicating that under such a tight bit budget a task-aware compression strategy can provide a noticeable advantage.} It is also seen that the MID closely tracks the RD benchmark, which also improves with $\mathsf{SNR}$, though we recall that it serves as a \emph{nonrealizable} reference, as it assumes an optimal Gaussian compression mechanism and an idealized reconstruction that cannot be implemented in practical systems.

We additionally report the corresponding bound-based performance for MID. In this case, the threshold is chosen by inverting the FA bound, so that the bound matches the target level $\epsilon_{\fa}$, and the MD bound is then evaluated at this threshold. The resulting curve is again valid but conservative, and the gap to the estimated performance curve is most pronounced in the mid-SNR range. Overall, these results demonstrate that MID rapidly improves with the SNR and provides a favorable reliability--SNR tradeoff in practical, nonasymptotic settings.

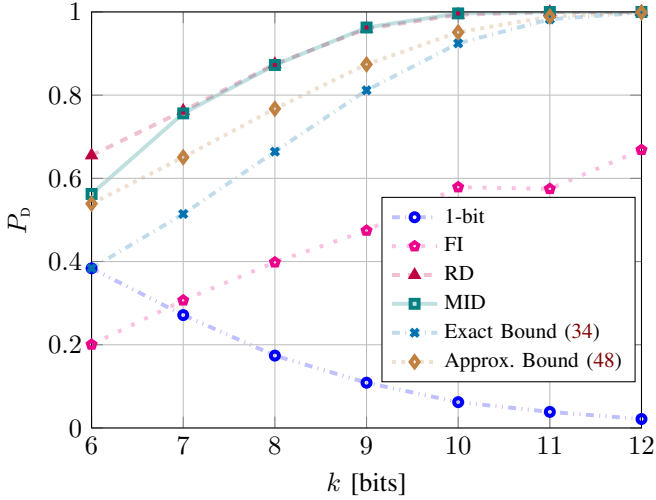
\begin{figure}[t]
    \centering
    
        \begin{tikzpicture}
            \begin{axis}[
                width=\linewidth,
                height=0.8*\linewidth,
                axis on top = false,
                grid=major,
                xlabel={$k$ [bits]},
                ylabel={$P_{\detc}$},
                xmin=6, xmax=12,
                ymin=0, ymax=1,
                xtick={6,7,8,9,10,11,12},
                ytick={0,0.2,0.4,0.6,0.8,1},
                legend pos=south east,
                legend cell align={left},
                legend style={font=\footnotesize, xshift=4pt, yshift=12.5pt},
                axis on top,
                domain=0:1,
            ]

            \addplot [
                dashdotdotted,
                color=blue,
                draw opacity=0.25,
                line width=1.5pt,
                mark=o,
                mark size=1.5pt,
                mark options={
                    solid,
                    draw opacity=1,
                    fill opacity=1,
                    fill=blue 
                  }
            ]
            table [x index=0, y index=1] {data/pd_vs_k_1b.dat};

            \addplot [
                loosely dotted,
                color=magenta,
                draw opacity=0.25,
                line width=1.5pt,
                mark=pentagon,
                mark size=1.5pt,
                mark options={
                    solid,
                    draw opacity=1,
                    fill opacity=1,
                    fill=magenta 
                  }
            ]
            table [x index=0, y index=1] {data/fig4_pd_vs_k_fowler_fair.dat};

            \addplot [
                dashed,
                color=purple,
                draw opacity=0.25,
                mark=triangle,
                mark size=1.5pt,
                line width=1.5pt,
                mark options={
                    solid,
                    draw opacity=1,
                    fill opacity=1,
                    fill=purple 
                  }
            ]
            table [x index=0, y index=1] {data/pd_vs_k_RD.dat};

            \addplot [
                color=teal,
                draw opacity=0.25,
                mark=square,
                mark size=1.5pt,
                line width=1.5pt,
                mark options={
                    solid,
                    draw opacity=1,
                    fill opacity=1,
                    fill=teal 
                  }
            ]
            table [x index=0, y index=1] {data/pd_vs_k_ext.dat};

            \addplot [
                dashdotted,
                color=uppboundcolor,
                draw opacity=0.25,
                mark=x,
                mark size=2pt,
                line width=1.5pt,
                mark options={
                    solid,
                    draw opacity=1,
                    fill opacity=1,
                    fill=uppboundcolor 
                  }
            ]
            table [x index=0, y index=1] {data/pd_vs_k_bound_exact.dat};

            \addplot [
                dotted,
                color=brown,
                draw opacity=0.25,
                mark=diamond,
                line width=1.5pt,
                mark options={
                    solid,
                    draw opacity=1,
                    fill opacity=1,
                    fill=brown 
                  }
            ]
            table [x index=0, y index=1] {data/pd_vs_k_bound_approx.dat};

            \legend{1-bit, FI, RD, MID, Exact Bound~\eqref{eq:Pmd-bound-u-ct}, Approx\addra{.}\delra{imated} Bound~\eqref{eq:Pmd-Laplace-approx-simplified}}
            \end{axis}
        \end{tikzpicture}\vspace{-0.2cm}

    \caption{Probability of detection versus $k$ for $N=2^k$~Nyquist samples, $\delta_m=\left\lfloor (N-1)/4 \right\rfloor$~sec, $\mathsf{SNR}_{\rndx}=3$~dB, $\mathsf{SNR}_{\rndy}=4$~dB and a fixed $\epsilon_{\fa}=10^{-3}$. Already at $k=7$, MID attains the theoretical RD performance, while considerably outperforming the 1-bit practical benchmark, which degrades as $\delta_m$ grows with $N$. Again, the exact and approximated upper bounds are conservative, but improve as $k$ increases and reflect the correct trend.}
    \label{fig:pd_vs_k}\vspace{-0.4cm}
\end{figure}

A complementary sweep over $k$, the communication budget in bits, is presented in Fig.~\ref{fig:pd_vs_k} for fixed (and this time, different) SNR levels $\mathsf{SNR}_{\rndx}=3$ dB and $\mathsf{SNR}_{\rndy}=4$ dB, and a growing (with $N$) maximum delay $\delta_m=\left\lfloor (N-1)/4 \right\rfloor$. The favorable behavior of our proposed method is evident: MID gains detection power rapidly with the message size, while closely tracking the theoretical (nonrealizable) RD benchmark. In contrast, the performance of the 1-bit scheme degrades as the delay uncertainty $\delta_{m}$ grows linearly with $N$\addra{, whereas FI improves, yet remains substantially weaker than MID}.

While our analysis in this paper focuses on the Gaussian signal model, we also evaluate the MID empirically in a representative non-Gaussian, heavy-tailed setting to assess its robustness outside the Gaussian realm. Specifically, we repeat the two simulation experiments in this section while keeping the same settings, but we replace the common source process $\rnds(t)$ by a bandlimited WSS process whose Nyquist samples are i.i.d.~Student's $t$ distributed with $\nu=5$ degrees of freedom (scaled to preserve the same variance, and hence the same nominal SNR definition), and an orthogonal frequency-division multiplexing (OFDM) signal with $N_{\text{\tiny FFT}}=2^k$ subcarriers, quadrature phase shift keying (QPSK) symbols placed on a fixed set of active tones, and a cyclic prefix of length $N_{\text{\tiny CP}}=N_{\text{\tiny FFT}}/4$, where Hermitian symmetry is imposed to yield a real-valued time-domain waveform. In both cases, the observation additive noises remain Gaussian, and we use the same thresholding procedure at the fixed design level $\epsilon_{\fa}$. 

\begin{figure}[t]
    \centering
    
        \begin{tikzpicture}
            \begin{axis}[
                width=\linewidth,
                height=0.8*\linewidth,
                axis on top = false,
                grid=major,
                xlabel={$k$ [bits]},
                ylabel={$P_{\detc}$},
                xmin=6, xmax=12,
                ymin=0, ymax=1,
                xtick={6,7,8,9,10,11,12},
                ytick={0,0.2,0.4,0.6,0.8,1},
                legend pos=north east,
                legend cell align={left},
                legend style={font=\footnotesize, xshift=4pt, yshift=-65pt},
                axis on top,
                domain=0:1,
            ]

            \addplot [
                dashdotdotted,
                color=blue,
                draw opacity=0.25,
                mark=square,
                mark size=1.5pt,
                line width=1.5pt,
                mark options={
                    solid,
                    draw opacity=1,
                    fill opacity=1,
                    fill=blue 
                  }
            ]
            table [x index=0, y index=1] {data/pd_vs_k_1b_students_t.dat};

            \addplot [
                dashdotdotted,
                color=blue,
                draw opacity=0.25,
                mark=o,
                mark size=1.5pt,
                line width=1.5pt,
                mark options={
                    solid,
                    draw opacity=1,
                    fill opacity=1,
                    fill=blue 
                  }
            ]
            table [x index=0, y index=1] {data/pd_vs_k_1b_OFDM.dat};

            \addplot [
                dashdotdotted,
                color=magenta,
                draw opacity=0.25,
                mark=square,
                mark size=1.5pt,
                line width=1.5pt,
                mark options={
                    solid,
                    draw opacity=1,
                    fill opacity=1,
                    fill=magenta 
                  }
            ]
            table [x index=0, y index=1] {data/fig6_pd_vs_k_fowler_fair_students_t.dat};

            \addplot [
                dashdotdotted,
                color=magenta,
                draw opacity=0.25,
                mark=o,
                mark size=1.5pt,
                line width=1.5pt,
                mark options={
                    solid,
                    draw opacity=1,
                    fill opacity=1,
                    fill=magenta 
                  }
            ]
            table [x index=0, y index=1] {data/fig6_pd_vs_k_fowler_fair_OFDM.dat};
            
            \addplot [
                color=teal,
                draw opacity=0.25,
                mark=square,
                mark size=1.5pt,
                line width=1.5pt,
                mark options={
                    solid,
                    draw opacity=1,
                    fill opacity=1,
                    fill=teal 
                  }
            ]
            table [x index=0, y index=1] {data/pd_vs_k_ext_students_t.dat};

            \addplot [
                color=teal,
                draw opacity=0.25,
                mark=o,
                mark size=1.5pt,
                line width=1.5pt,
                mark options={
                    solid,
                    draw opacity=1,
                    fill opacity=1,
                    fill=teal 
                  }
            ]
            table [x index=0, y index=1] {data/pd_vs_k_ext_OFDM.dat};

            \legend{{1-bit, Student's $t$}, {1-bit, OFDM}, {FI, Student's $t$}, {FI, OFDM}, {MID, Student's $t$}, {MID, OFDM}}
            \end{axis}
        \end{tikzpicture}\vspace{-0.2cm}

    \caption{Probability of detection versus $k$ for the same setting detailed in the caption of Fig.~\ref{fig:pd_vs_k}, but with non-Gaussian sources.}
    \label{fig:pd_vs_k_nonGaussian}\vspace{-0.4cm}
\end{figure}
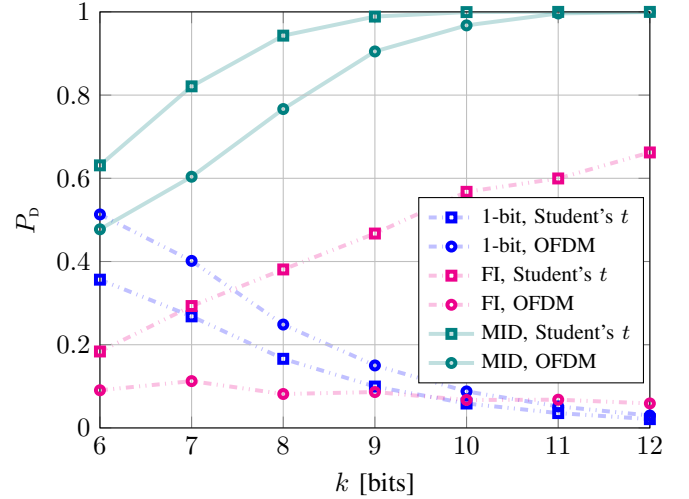

The resulting detection curves, presented in Figs.~\ref{fig:pd_vs_snr_nonGaussian} and~\ref{fig:pd_vs_k_nonGaussian}, exhibit the same qualitative trends observed under Gaussianity: the MID continues to improve markedly with increasing SNR and bit budget, and retains a clear advantage over the competing $1$-bit\addra{ and FI} realizable baseline\addra{s}. We have qualitatively reproduced the same behavior under additional non-Gaussian source distributions (not reported here for brevity). These results indicate that the proposed extremum-based mechanism is not intrinsically tied to Gaussianity and can remain effective even under heavy-tailed signal statistics. A rigorous theoretical extension of the nonasymptotic guarantees to non-Gaussian and temporally-correlated signals, while outside the scope of the present work, will be pursued in future work.

\begin{figure}[t]
    \centering
    
        \begin{tikzpicture}
            \begin{axis}[
                width=\linewidth,
                height=0.8*\linewidth,
                axis on top = false,
                grid=major,
                xlabel={$\mathsf{SNR}$ [dB]},
                ylabel={$P_{\detc}$},
                xmin=-10, xmax=10,
                ymin=0, ymax=1,
                xtick={-10,-8,-6,-4,-2,0,2,4,6,8,10},
                ytick={0,0.2,0.4,0.6,0.8,1},
                legend pos=north west,
                legend cell align={left},
                legend style={font=\footnotesize},
                axis on top,
                domain=0:1,
            ]

            \addplot [
                dashdotdotted,
                color=blue,
                mark=square,
                mark size=1.5pt,
                mark repeat=2,
                line width=1.5pt,
                mark options={
                    solid,
                    draw opacity=1,
                    fill opacity=1,
                    fill=blue 
                  }
            ]
            table [x index=0, y index=1] {data/pd_vs_snr_1b_students_t.dat};

            \addplot [
                dashdotdotted,
                color=blue,
                mark=o,
                mark repeat=2,
                mark size=1.5pt,
                line width=1.5pt,
                mark options={
                    solid,
                    draw opacity=1,
                    fill opacity=1,
                    fill=blue 
                  }
            ]
            table [x index=0, y index=1] {data/pd_vs_snr_1b_OFDM.dat};

            \addplot [
                loosely dotted,
                color=magenta,
                mark=square,
                mark repeat=2,
                mark size=1.5pt,
                line width=1.5pt,
                mark options={
                    solid,
                    draw opacity=1,
                    fill opacity=1,
                    fill=magenta 
                  }
            ]
            table [x index=0, y index=1] {data/fig5_pd_vs_snr_fowler_fair_students_t.dat};

            \addplot [
                loosely dotted,
                color=magenta,
                mark=o,
                mark repeat=2,
                mark size=1.5pt,
                line width=1.5pt,
                mark options={
                    solid,
                    draw opacity=1,
                    fill opacity=1,
                    fill=magenta 
                  }
            ]
            table [x index=0, y index=1] {data/fig5_pd_vs_snr_fowler_fair_OFDM.dat};
            
            \addplot [
                color=teal,
                mark=square,
                mark repeat=2,
                mark size=1.5pt,
                line width=1.5pt,
                mark options={
                    solid,
                    draw opacity=1,
                    fill opacity=1,
                    fill=teal 
                  }
            ]
            table [x index=0, y index=1] {data/pd_vs_snr_ext_students_t.dat};

            \addplot [
                color=teal,
                mark=o,
                mark repeat=2,
                mark size=1.5pt,
                line width=1.5pt,
                mark options={
                    solid,
                    draw opacity=1,
                    fill opacity=1,
                    fill=teal 
                  }
            ]
            table [x index=0, y index=1] {data/pd_vs_snr_ext_OFDM.dat};

            \legend{{1-bit, Student's $t$}, {1-bit, OFDM}, {FI, Student's $t$}, {FI, OFDM}, {MID, Student's $t$}, {MID, OFDM}}
            \end{axis}
        \end{tikzpicture}\vspace{-0.2cm}

    \caption{Probability of detection versus SNR for the same setting detailed in the caption of Fig.~\ref{fig:pd_vs_snr}, but with non-Gaussian sources.}
    \label{fig:pd_vs_snr_nonGaussian}\vspace{-0.3cm}
\end{figure}
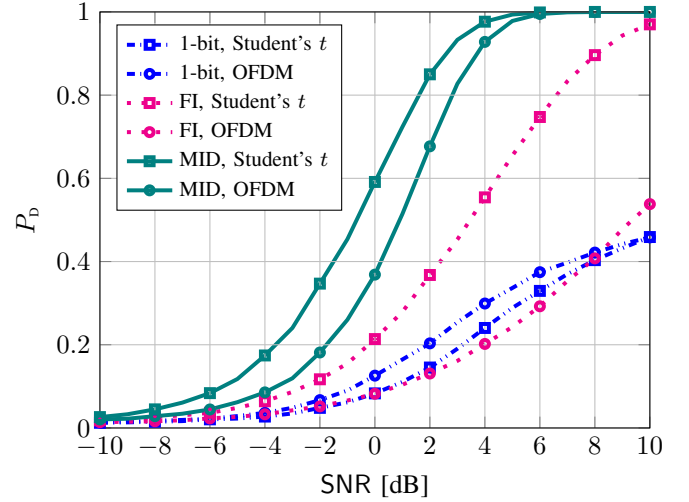

\vspace{-0.2cm}
\addraR{\subsection{Robustness to Multipath Propagation}\label{subsec:multipath_robustness}}
\addraR{We next examine the robustness of MID to a multipath propagation model. This simulation experiment is motivated by the fact that the analysis above is developed for the single-delay model \eqref{eq:h1}, whereas in some propagation environments the received signal may contain several delayed replicas of the same underlying source. Specifically, under $\setH_1$ we replace the single delayed component at the decoder by}
\begin{equation}
\addraR{\rndy(t)=\sum_{q=0}^{Q-1} a_q \rnds(t-\delta_q)+\rndz_2(t), \qquad \sum_{q=0}^{Q-1}a_q^2=1,}
\label{eq:multipath_model}
\end{equation}
\addraR{where $\delta_q\in\Delta$ are unknown deterministic path delays and the normalization keeps the received source power fixed. Importantly, the detector itself is not modified: the encoder still transmits only the single maximum index $\rndj$, and the decoder still uses the same statistic $\rndt_{\midet}$ \eqref{eq:test_stat_ct}. Thus, this experiment tests whether the proposed single-index mechanism is intrinsically brittle to multipath, or whether it remains effective without any multipath-specific modification. All empirical results in this experiment were obtained based on $10^5$ independent trials.}

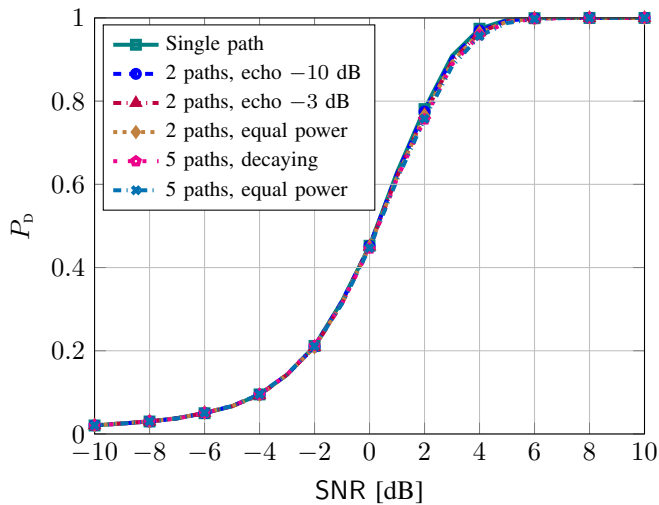
\begin{figure}[t]
    \centering
        \begin{tikzpicture}
            \begin{axis}[
                width=\linewidth,
                height=0.8*\linewidth,
                axis on top = false,
                grid=major,
                xlabel={\addraR{$\mathsf{SNR}$ [dB]}},
                ylabel={\addraR{$P_{\detc}$}},
                xmin=-10, xmax=10,
                ymin=0, ymax=1,
                xtick={-10,-8,-6,-4,-2,0,2,4,6,8,10},
                ytick={0,0.2,0.4,0.6,0.8,1},
                legend pos=north west,
                legend cell align={left},
                legend style={font=\footnotesize,
                 xshift=-1.1mm,
                 yshift=0.8mm},
                axis on top,
                domain=0:1,
            ]

            \addplot [
                color=teal,
                mark=square,
                mark repeat=2,
                mark size=1.5pt,
                line width=1.5pt,
                mark options={solid, draw opacity=1, fill opacity=1, fill=teal}
            ]
            table [x index=0, y index=1] {data/pd_vs_snr_multipath_single.dat};

            \addplot [
                dashed,
                color=blue,
                mark=o,
                mark repeat=2,
                mark size=1.5pt,
                line width=1.5pt,
                mark options={solid, draw opacity=1, fill opacity=1, fill=blue}
            ]
            table [x index=0, y index=1] {data/pd_vs_snr_multipath_two_echo_m10db.dat};

            \addplot [
                dashdotted,
                color=purple,
                mark=triangle,
                mark repeat=2,
                mark size=1.5pt,
                line width=1.5pt,
                mark options={solid, draw opacity=1, fill opacity=1, fill=purple}
            ]
            table [x index=0, y index=1] {data/pd_vs_snr_multipath_two_echo_m3db.dat};

            \addplot [
                dotted,
                color=brown,
                mark=diamond,
                mark repeat=2,
                mark size=1.5pt,
                line width=1.5pt,
                mark options={solid, draw opacity=1, fill opacity=1, fill=brown}
            ]
            table [x index=0, y index=1] {data/pd_vs_snr_multipath_two_equal.dat};

            \addplot [
                loosely dotted,
                color=magenta,
                mark=pentagon,
                mark repeat=2,
                mark size=1.5pt,
                line width=1.5pt,
                mark options={solid, draw opacity=1, fill opacity=1, fill=magenta}
            ]
            table [x index=0, y index=1] {data/pd_vs_snr_multipath_five_decay.dat};

            \addplot [
                dashdotdotted,
                color=uppboundcolor,
                mark=x,
                mark repeat=2,
                mark size=2pt,
                line width=1.5pt,
                mark options={solid, draw opacity=1, fill opacity=1, fill=uppboundcolor}
            ]
            table [x index=0, y index=1] {data/pd_vs_snr_multipath_five_equal.dat};

            \legend{\addraR{Single path}, \addraR{2 paths, echo $-10$ dB}, \addraR{2 paths, echo $-3$ dB}, \addraR{2 paths, equal power}, \addraR{5 paths, decaying}, \addraR{5 paths, equal power}}
            \end{axis}
        \end{tikzpicture}\vspace{-0.2cm}

    \caption{\addraR{Probability of detection versus SNR for the (unchanged) MID detector under the multipath model \eqref{eq:multipath_model}, with $k=7$ bits, $N=128$ Nyquist samples, $\delta_m=60$ sec and a fixed $\epsilon_{\fa}=10^{-2}$. The single-path curve corresponds to the original model and is shown as a reference; the decaying five-path profile uses relative powers $0,-3,-6,-9,-12$ dB.}}
    \label{fig:pd_vs_snr_multipath}\vspace{-0.45cm}
\end{figure}

\addraR{Fig.~\ref{fig:pd_vs_snr_multipath} shows the resulting detection performance for the considered two-path and five-path profiles, together with the single-path reference. The curves indicate only moderate-to-negligible degradation relative to the single-path case. In particular, the detector continues to improve rapidly with SNR and reaches reliable detection at effectively the same SNR range as in the single-path case. This behavior is consistent with the structure of the statistic: although only one encoder-side anchor is transmitted, the decoder-side maximization over the delay can align this anchor with any sufficiently strong delayed replica of the source. Thus, while the decoder may align the encoder-side anchor with a non-line-of-sight component, multipath does not necessarily invalidate the proposed detector.}

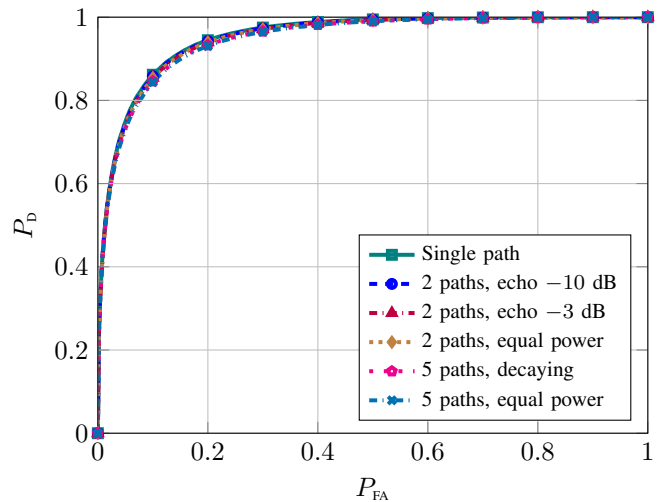
\begin{figure}[t]
    \centering
        \begin{tikzpicture}
            \begin{axis}[
                width=\linewidth,
                height=0.8*\linewidth,
                axis on top = false,
                grid=major,
                xlabel={\addraR{$P_{\fa}$}},
                ylabel={\addraR{$P_{\detc}$}},
                xmin=0, xmax=1,
                ymin=0, ymax=1,
                xtick={0,0.2,0.4,0.6,0.8,1},
                ytick={0,0.2,0.4,0.6,0.8,1},
                legend pos=south east,
                legend cell align={left},
                legend style={font=\footnotesize},
                axis on top,
                domain=0:1,
            ]

            \addplot [
                color=teal,
                mark=square,
                mark repeat=50,
                mark size=1.5pt,
                line width=1.5pt,
                mark options={solid, draw opacity=1, fill opacity=1, fill=teal}
            ]
            table [x index=0, y index=1] {data/ROC_multipath_single.dat};

            \addplot [
                dashed,
                color=blue,
                mark=o,
                mark repeat=50,
                mark size=1.5pt,
                line width=1.5pt,
                mark options={solid, draw opacity=1, fill opacity=1, fill=blue}
            ]
            table [x index=0, y index=1] {data/ROC_multipath_two_echo_m10db.dat};

            \addplot [
                dashdotted,
                color=purple,
                mark=triangle,
                mark repeat=50,
                mark size=1.5pt,
                line width=1.5pt,
                mark options={solid, draw opacity=1, fill opacity=1, fill=purple}
            ]
            table [x index=0, y index=1] {data/ROC_multipath_two_echo_m3db.dat};

            \addplot [
                dotted,
                color=brown,
                mark=diamond,
                mark repeat=50,
                mark size=1.5pt,
                line width=1.5pt,
                mark options={solid, draw opacity=1, fill opacity=1, fill=brown}
            ]
            table [x index=0, y index=1] {data/ROC_multipath_two_equal.dat};

            \addplot [
                loosely dotted,
                color=magenta,
                mark=pentagon,
                mark repeat=50,
                mark size=1.5pt,
                line width=1.5pt,
                mark options={solid, draw opacity=1, fill opacity=1, fill=magenta}
            ]
            table [x index=0, y index=1] {data/ROC_multipath_five_decay.dat};

            \addplot [
                dashdotdotted,
                color=uppboundcolor,
                mark=x,
                mark repeat=50,
                mark size=2pt,
                line width=1.5pt,
                mark options={solid, draw opacity=1, fill opacity=1, fill=uppboundcolor}
            ]
            table [x index=0, y index=1] {data/ROC_multipath_five_equal.dat};

            \legend{\addraR{Single path}, \addraR{2 paths, echo $-10$ dB}, \addraR{2 paths, echo $-3$ dB}, \addraR{2 paths, equal power}, \addraR{5 paths, decaying}, \addraR{5 paths, equal power}}
            \end{axis}
        \end{tikzpicture}\vspace{-0.2cm}

    \caption{\addraR{ROC curves for the same multipath profiles considered in Fig.~\ref{fig:pd_vs_snr_multipath}, with $k=7$ bits, $N=128$ samples, $\delta_m=60$ sec and $\mathsf{SNR}_{\rndx}=\mathsf{SNR}_{\rndy}=0$ dB.}}
    \label{fig:roc_multipath}\vspace{-0.45cm}
\end{figure}

\addraR{The ROC curves in Fig.~\ref{fig:roc_multipath} show the same qualitative behavior across the full range of operating points, indicating that the conclusion is not tied to the particular choice $\epsilon_{\fa}=10^{-2}$. At the same time, the experiment should not be interpreted as a complete multipath theory. If the path gains and delays create a highly diffuse or destructive effective correlation profile, the aligned component may be substantially reduced, and extensions based on several extrema or more structured candidate alignments may become useful.}

\section{Concluding Remarks}\label{sec:conclusion}
This work studies the problem of joint compression and detection in a two-sensor time-delay propagation model under an explicit $k$-bit communication constraint. We propose a low-complexity extremum-based encoder--detector scheme that leverages the delay structure while respecting a strict bit budget. For the proposed scheme, we derive nonasymptotic upper bounds on the FA and MD probabilities, with a tractable, asymptotically accurate approximation for the latter. Simulations demonstrate the resulting trade-off between communication budget and detection performance, as well as the superiority of our scheme relative to both practical and theoretical, more conventional baselines.

Still, many important questions remain to be addressed. First, on the fundamental side, a tight \emph{converse} (nontrivial lower bound) on the achievable error probability under $k$-bit constraints is still missing. \addra{Deriving such a converse appears considerably more challenging here than in standard rate--distortion or related compression--inference settings, due to the combination of continuous-time observations, finite-rate encoding, and detection under an unknown continuous-valued time delay. }Likewise, there is room for tighter upper bounds, e.g., by sharpening some of the intermediate inequalities or by exploiting more structure in the extremum statistics.

\delraR{\addra{On the modeling side, another practically important direction is extension beyond the single-delay model considered here, e.g., to multipath settings. In such cases, the encoder--detector architecture would likely need to account for multiple candidate alignments rather than a single dominant one, for example via several prominent extrema or a more structured local search. While some of the intuition underlying the current model may extend naturally to multipath settings, the corresponding theory would most likely become substantially more involved, and practical schemes would likely incur significantly higher computational complexity.}}

\addraR{The multipath simulations in Section~\ref{subsec:multipath_robustness} further suggest that the proposed MID detector has a meaningful degree of robustness beyond the single-delay model analyzed theoretically in this paper. In particular, even without modifying the encoder or decoder, the single transmitted extremum index remains useful in representative two-path and five-path settings, since the decoder-side maximization over the delay can align the encoder-side anchor with a sufficiently strong delayed replica. Thus, moderate multipath does not appear to invalidate the proposed mechanism. A complete theory for arbitrary multipath channels, as well as multipath-specific designs based on several extrema or more structured candidate alignments, remains an important direction for future work, especially for highly diffuse or destructive propagation profiles.}

\addra{Yet another natural extension is to go beyond transmitting only a single extremum index, and instead use the available bit budget to convey richer candidate information, such as several prominent peak locations or a combination of coarse location and quantized amplitude. The resulting tradeoffs appear to be nontrivial and may depend on the operating regime. In particular, even the comparison with the proposed maximum-only scheme is not entirely straightforward, since one may wish to hold fixed the total bit budget, the observation length, or both jointly. More broadly, one must balance the reliability of the single most informative candidate against the benefit of conveying additional structural information. Investigating such extensions requires a systematic theoretical treatment.}

It is also natural to ask how our scheme and analysis extend beyond the present Gaussian setting\addra{. On the one hand, the non-Gaussian simulation experiments in Section~\ref{sec:simulation} suggest that the proposed scheme may remain effective beyond the Gaussian model. On the other hand, the current analysis relies on tools specific to Gaussian processes, including Rice-formula arguments, Gaussian MMSE decompositions, and Gaussian extreme-value approximations. Thus, important extensions include}\delra{: to} non-Gaussian random processes (e.g., impulsive or heavy-tailed noise), \delra{to }partially deterministic or deterministic-but-unknown signals, \delra{or to}\addra{and} models with structured priors on the delay or channel. Another key direction is generalization from two sensors to larger networks, where questions of encoder cooperation, bit allocation across sensors, and possibly multi-hop or hierarchical architectures become central. Additional extensions include robustness to model mismatch, joint \delra{time-delay estimation}\addra{TDE} and detection under a unified compression strategy, possibly with an observation time that grows slower with $k$, and learning-based encoders that adapt the extremum selection rule or quantization to observed data.

Overall, the present work should be viewed as a building block toward a broader theory and practice of joint compression and inference in distributed sensing systems under hard communication budgets. The extremum-based architecture and the accompanying nonasymptotic analysis provide a concrete, analytically tractable starting point on which more sophisticated schemes and tighter performance limits can be developed. We expect that pursuing the open questions outlined above will lead to a richer understanding of the fundamental limits and to practical designs for future large-scale sensing networks operating under stringent communication constraints.

\appendices
\section{Derivation of the GLRT}\label{app:GLRT}
In this appendix we derive the GLRT \eqref{eq:ct_glrt_corr} with unlimited communication, i.e., based on the signals $\{\rndx(t),\rndy(t)\}_{t\in(0,T)}$.

Fix an admissible time-delay $\tau\in\Delta$ and define the $\tau$-aligned second signal
\begin{equation}
    \rndy_\tau(t) \triangleq \rndy(t+\tau),\qquad t\in(0,T).
\end{equation}
We treat the pair $(\rndx(t),\rndy_\tau(t))$ as a random element of the
signal space $L^2(0,T)\times L^2(0,T)$ equipped with the standard inner
product
\begin{align}
    & {\langle} \tp{[x_1(t)\; y_1(t)]},\tp{[x_2(t)\; y_2(t)]}\rangle \\
    &\qquad {\triangleq} \int_0^T x_1(t)x_2(t)\,{\rm d}t
              + \int_0^T y_1(t)y_2(t)\,{\rm d}t.
\end{align}

Let $\{\varphi_m(t)\}_{m=1}^\infty$ be any complete orthonormal basis of
$L^2(0,T)$; for example, a Fourier series basis on $(0,T)$. For a fixed integer $M\ge 1$, consider
the first $M$ basis functions and define the finite set of projection
coefficients
\begin{align}
    \rndx_m &\triangleq \int_0^T \rndx(t)\,\varphi_m(t)\,{\rm d}t,
    \label{eq:def_xm}\\
    \rndy_{\tau,m} &\triangleq \int_0^T \rndy_\tau(t)\,\varphi_m(t)\,{\rm d}t
    = \int_0^T \rndy(t+\tau)\,\varphi_m(t)\,{\rm d}t,
    \label{eq:def_ym}
\end{align}
for $m=1,\ldots,M$. Under either hypothesis, the random vector
\begin{equation}
    \rvecz^{(M)}_\tau
    \triangleq \tp{[\rndx_1\;\cdots\;\rndx_M\;\rndy_{\tau,1}\;\cdots\;\rndy_{\tau,M}]}
    \in\reals^{2M\times 1}
\end{equation}
is zero-mean Gaussian, since it is formed by linear functionals of the jointly
Gaussian processes $\rndx(\cdot)$ and $\rndy(\cdot)$.

The covariance matrix of $\rvecz^{(M)}_\tau$ under $\setH_i$, $i\in\{0,1\}$,
is completely determined by the second-order statistics of $\rndx(t)$ and
$\rndy_\tau(t)$. In particular, for each $m$,
\begin{align}
    \Exop[\rndx_m^2]
    &= \int_0^T\!\!\int_0^T
         R_{\rndx\rndx}^{(i)}(t-u)\,\varphi_m(t)\varphi_m(u)\,
       {\rm d}u\,{\rm d}t,
    \label{eq:var_xm}\\
    \Exop[\rndy_{\tau,m}^2]
    &= \int_0^T\!\!\int_0^T
         R_{\rndy\rndy}^{(i)}(t-u)\,\varphi_m(t)\varphi_m(u)\,
       {\rm d}u\,{\rm d}t,
    \label{eq:var_ym}\\
    \Exop[\rndx_m \rndy_{\tau,m}]
    &= \int_0^T\!\!\int_0^T
         R_{\rndx\rndy}^{(i)}(t-u+\tau)\,\varphi_m(t)\varphi_m(u)\,
       {\rm d}u\,{\rm d}t.
    \label{eq:cov_xm_ym}
\end{align}
Here $R_{\rndx\rndx}^{(i)}$, $R_{\rndy\rndy}^{(i)}$ and $R_{\rndx\rndy}^{(i)}$
denote the autocorrelation and cross-correlation functions under $\setH_i$.
From \eqref{eq:h0}--\eqref{eq:h1} and~\eqref{eq:crosscorr} we have:
\begin{itemize}
    \item Under $\setH_0$, $\rndx(t)$ and $\rndy_\tau(t)$ are uncorrelated, so
    \begin{equation}
        R_{\rndx\rndy}^{(0)}(t-u) = 0
        \;\Longrightarrow\;
        \Exop[\rndx_m\rndy_{\tau,m}] = 0,\;\forall m\in\naturals.
    \end{equation}
    Hence, the covariance matrix of $\rvecz^{(M)}_\tau$ under $\setH_0$ is
    \emph{block-diagonal}.
    \item Under $\setH_1$, $\rndx(t)$ and $\rndy_\tau(t)$ share the common
    source $\rnds(t)$, and their cross-correlation is
    \begin{equation}
        R_{\rndx\rndy}^{(1)}(t-u) = R_{\rnds\rnds}(t-u-\tau+\delta),
    \end{equation}
    so that
    \begin{equation}
        \Exop[\rndx_m\rndy_{\tau,m}]
        = \int_0^T\!\!\int_0^T
            R_{\rnds\rnds}(t-u-\tau+\delta)\,\varphi_m(t)\varphi_m(u)\,
          {\rm d}u\,{\rm d}t.
    \end{equation}
    In particular, the cross-covariance block of $\rvecz^{(M)}_\tau$ under
    $\setH_1$ is generally nonzero.
\end{itemize}

Thus, for each finite $M$, the observed vector $\rvecz^{(M)}_\tau$ is a
zero-mean Gaussian vector whose covariance matrix under $\setH_0$ is
block-diagonal (no correlation between $\{\rndx_m\}$ and
$\{\rndy_{\tau,m}\}$), and under $\setH_1$ has a nonzero cross-covariance block.
We now derive the GLRT for this finite-dimensional Gaussian problem.

For each $M$ and time-delay $\tau$, denote by $\matSigma_{0}^{(M)}\in\reals^{2M\times 2M}$ and
$\matSigma_{1}^{(M)}(\tau)\in\reals^{2M\times 2M}$ the covariance matrices of $\rvecz^{(M)}_\tau$
under $\setH_0$ and $\setH_1$, respectively. Recall that the variances and the correlation
parameters as unknown nuisance parameters subject only to the structural
constraints:
\begin{itemize}
    \item under $\setH_0$, $\matSigma_{0}^{(M)}$ is block-diagonal;
    \item under $\setH_1$, $\matSigma_{1}^{(M)}(\tau)$ has a free off-diagonal
    block.
\end{itemize}

Let $p_i(\rvecz^{(M)}_\tau\mid\matSigma_i^{(M)})$ denote the $2M$-variate Gaussian density under $\setH_i$. The GLRT based on the first $M$ projections and the time-delay $\tau$ compares the statistic $\Lambda_M(\tau)$ to a threshold, where
\begin{equation}
    \Lambda_M(\tau)
    \triangleq
    \frac{\displaystyle\sup_{\matSigma_1^{(M)}}
       p_1\left(\rvecz^{(M)}_\tau\mid\matSigma_1^{(M)}\right)}
         {\displaystyle\sup_{\matSigma_0^{(M)}\in\mathcal{B}_M}
      p_0\left(\rvecz^{(M)}_\tau\mid\matSigma_0^{(M)}\right)},
\end{equation}
and here $\mathcal{B}_M$ is the set of $2M\times 2M$ block-diagonal positive-definite
covariance matrices. Standard algebra shows that:
\begin{itemize}
    \item the maximum likelihood estimate (MLE) of the covariance under $\setH_1$ is the empirical covariance of
    $\rvecz^{(M)}_\tau$ with full off-diagonal blocks;
    \item the MLE of the covariance under $\setH_0$ is obtained from the empirical covariance by zeroing the off-diagonal cross-block.
\end{itemize}

Denote the empirical covariances as
\begin{align}
    \widehat{\sigma}_{\rndx,M}^2 &\triangleq
    \frac{1}{M}\sum_{m=1}^M \rndx_m^2,\\
    \widehat{\sigma}_{\rndy,M}^2 &\triangleq
    \frac{1}{M}\sum_{m=1}^M \rndy_{\tau,m}^2,\\
    \widehat{\sigma}_{\rndx\rndy,M}(\tau) &\triangleq
    \frac{1}{M}\sum_{m=1}^M \rndx_m \rndy_{\tau,m}.
\end{align}
Then the empirical covariance matrices under $\setH_0$ and $\setH_1$ are
\begin{equation}
    \widehat{\matSigma}_0^{(M)}
    =
    \begin{bmatrix}
        \widehat{\sigma}_{\rndx,M}^2 \matI_M & 0\\
        0 & \widehat{\sigma}_{\rndy,M}^2 \matI_M
    \end{bmatrix},
\end{equation}
and
\begin{equation}
    \widehat{\matSigma}_1^{(M)}(\tau)
    =
    \begin{bmatrix}
        \widehat{\sigma}_{\rndx,M}^2 \matI_M
        & \widehat{\sigma}_{\rndx\rndy,M}(\tau) \matI_M\\
        \widehat{\sigma}_{\rndx\rndy,M}(\tau) \matI_M
        & \widehat{\sigma}_{\rndy,M}^2 \matI_M
    \end{bmatrix},
\end{equation}
respectively, where $\matI_M$ is the $M\times M$ identity matrix. Substituting
these into the Gaussian likelihoods and simplifying, we obtain
\begin{equation}
    \Lambda_M(\tau)
    = \left(
        \frac{\det\widehat{\matSigma}_0^{(M)}}
             {\det\widehat{\matSigma}_1^{(M)}(\tau)}
      \right)^{\!1/2}
    = \left(
        \frac{1}{1-\widehat{\rho}_M^2(\tau)}
      \right)^{\!M/2},
    \label{eq:LambdaM}
\end{equation}
where
\begin{equation}
    \widehat{\rho}_M(\tau) \triangleq \frac{\widehat{\sigma}_{\rndx\rndy,M}(\tau)}{
    \sqrt{\widehat{\sigma}_{\rndx,M}^2\,\widehat{\sigma}_{\rndy,M}^2}}.
    \label{eq:rhoM}
\end{equation}
Thus, for each fixed $M$ and $\tau$, the generalized likelihood ratio
$\Lambda_M(\tau)$ is a strictly increasing function of
$\widehat{\rho}_M^2(\tau)$. Therefore, the GLRT based on the first $M$ projections maximizes $\Lambda_M(\tau)$ over
$\tau\in\Delta$ and compares it to a threshold; equivalently, it compares
$\max_{\tau\in\Delta}\widehat{\rho}_M(\tau)$ to a threshold.

We now pass from the finite-dimensional statistic $\widehat{\rho}_M(\tau)$ to
its continuous-time counterpart. Using the definitions~\eqref{eq:def_xm} and
\eqref{eq:def_ym}, the numerator in~\eqref{eq:rhoM} can be written as
\begin{align}
    &\sum_{m=1}^M \rndx_m \rndy_{\tau,m}\\
    &= \sum_{m=1}^M
       \left(\int_0^T \rndx(t)\varphi_m(t)\,{\rm d}t\right)
       \left(\int_0^T \rndy(t'+\tau)\varphi_m(t')\,{\rm d}t'\right)\\
    &= \int_0^T\!\!\int_0^T
       \rndx(t)\rndy(t'+\tau)
       \left(\sum_{m=1}^M \varphi_m(t)\varphi_m(t')\right)
       {\rm d}t'\,{\rm d}t.
\end{align}
Similarly,
\begin{align}
    \sum_{m=1}^M \rndx_m^2
    &= \int_0^T\!\!\int_0^T
       \rndx(t)\rndx(t')
       \left(\sum_{m=1}^M \varphi_m(t)\varphi_m(t')\right)
       {\rm d}t'\,{\rm d}t,\\
    \sum_{m=1}^M \rndy_{\tau,m}^2
    &= \int_0^T\!\!\int_0^T
       \rndy_{\tau}(t)\rndy_{\tau}(t')
       \left(\sum_{m=1}^M \varphi_m(t)\varphi_m(t')\right)
       {\rm d}t'\,{\rm d}t.
\end{align}

Since $\{\varphi_m\}_{m=1}^\infty$ is an orthonormal basis of $L^2(0,T)$, the
kernel
\begin{equation}
    K_M(t,t') \triangleq \sum_{m=1}^M \varphi_m(t)\varphi_m(t')
\end{equation}
converges in the $L^2$ sense (and pointwise almost everywhere) to the Dirac
kernel on $(0,T)$ in the usual reproducing sense:
\begin{equation}
    \lim_{M\to\infty} \int_0^T K_M(t,t')f(t')\,{\rm d}t' = f(t)
    \quad \forall f\in L^2(0,T).
\end{equation}
Applying this to $f(t') = \rndx(t')$ and $f(t') = \rndy(t'+\tau)$, we obtain, almost surely,
\begin{align}
    \lim_{M\to\infty} \sum_{m=1}^M \rndx_m^2
    &= \int_0^T \rndx^2(t)\,{\rm d}t,\\
    \lim_{M\to\infty} \sum_{m=1}^M \rndy_{\tau,m}^2
    &= \int_0^T \rndy^2(t+\tau)\,{\rm d}t.\label{eq:defofyenergywithdelta}
\end{align}
Moreover, by continuity of the $L^2$ inner product, we have, almost surely,
\begin{align}
    \lim_{M\to\infty} \sum_{m=1}^M \rndx_m \rndy_{\tau,m}
    &= \int_0^T \rndx(t)\rndy(t+\tau)\,{\rm d}t.
\end{align}
Therefore, as $M\to\infty$,
\begin{equation}
    \widehat{\rho}_M(\tau)
    \;\longrightarrow\;
    \widehat{\rho}_{\uc}(\tau)
    =
    \frac{\displaystyle\int_0^T \rndx(t)\rndy(t+\tau)\,{\rm d}t}{
    \sqrt{\displaystyle\int_0^T \rndx^2(t)\,{\rm d}t
         \displaystyle\int_0^T \rndy^2(t+\tau)\,{\rm d}t}},
    \label{eq:rho_ct_def}
\end{equation}
and, from~\eqref{eq:LambdaM}, the corresponding log-likelihood ratios
$\log\Lambda_M(\tau)$ converge to a strictly increasing function of
$\widehat{\rho}_{\uc}^2(\tau)$.

By definition, the GLRT for the continuous-time problem is obtained as the
limit, as $M\to\infty$, of the GLRTs based on the finite-dimensional
projections, since the set of all projections $\{\rndx_m,\rndy_{\tau,m}\}$
captures all the information contained in $(\rndx(t),\rndy(t+\tau))$ over
$(0,T)$. Hence, the continuous-time GLRT statistic for a fixed time-delay
$\tau$ is a strictly increasing function of $\widehat{\rho}_{\uc}^2(\tau)$, where
$\widehat{\rho}_{\uc}(\tau)$ is defined by~\eqref{eq:normalizedcrosscorr}.

Finally, we must account for the unknown delay $\tau\in\Delta$. The GLRT
maximizes the likelihood ratio over $\tau$,
\begin{equation}
    \Lambda_{\GLRT}
    \triangleq \sup_{\tau\in\Delta} \Lambda(\tau),
\end{equation}
and compares it to a threshold. Since $\Lambda(\tau)$ is a strictly increasing
function of $\widehat{\rho}_{\uc}^2(\tau)$, this GLRT is equivalent to the
correlation test
\begin{equation}
    \max_{\tau\in\Delta} \widehat{\rho}_{\uc}(\tau)
    \;\mathop{\gtrless}_{\setH_0}^{\setH_1}\; \gamma_{\uc},
    \label{eq:GLRT_ct_final}
\end{equation}
for a threshold $\gamma_{\uc}\in\reals$ chosen to meet a desired FA tolerance.

In words, in the unlimited-communication setting, the GLRT computes, for each $\tau\in\Delta$, the normalized empirical
cross-correlation $\widehat{\rho}_{\uc}(\tau)$, takes the maximum over the delay interval $\Delta$, and compares this maximum to a threshold. 

\section{Unbiasedness of the Estimator \eqref{eq:ccdefcommconst}}\label{app:unbiasedness}
We would like to show that $\Exop\left[ \widehat{\rho}_{\lc}(\tau) \mid \setH\right] = \rho(\tau)$, where
\begin{align}
    \rho(\tau) &\triangleq \frac{\Exop\left[ \rndy(t)\rndx(t-\tau) \mid \setH \right]}{\sqrt{\Exop\left[\rndy^2(t) \mid \setH \right]\Exop\left[ \rndx^2(t) \mid \setH \right]}} \\
    &= \begin{cases}
        0, & \setH=\setH_0 \\
        \frac{R_{\rnds\rnds}(\tau-\delta)}{\sigma_{\rndy}\sigma_{\rndx}}, & \setH=\setH_1
    \end{cases},\label{eq:unbiasedcondition}
\end{align}
and $\sigma_{\rndx}^2\triangleq \sigma_{\rnds}^2+\sigma^2_1$ and $\sigma_{\rndy}^2\triangleq \sigma_{\rnds}^2+\sigma^2_2$. For $\setH=\setH_0$, this is trivial, since in this case $\rndy(t)=\rndz_2(t)$ is independent of $\rndj$ for all $t\in\reals$, and therefore
\begin{align}
    \Exop\left[ \widehat{\rho}_{\lc}(\tau) \mid \setH_0\right] &= \frac{\sigma_{\rndx}}{\sigma_{\rndy}}\cdot \frac{\Exop[\rndy(t_{\rndj} + \tau)\mid \setH_0]}{\Exop\left[\rndx[\rndj] \mid \setH_1\right]} \\
    &=\frac{\sigma_{\rndx}}{\sigma_{\rndy}}\cdot \frac{\Exop[\rndz_2(t_{\rndj} + \tau)]}{\Exop\left[\rndx[\rndj] \mid \setH_1\right]}\\
    &=\frac{\sigma_{\rndx}}{\sigma_{\rndy}}\cdot \frac{\Exop[\rndz_2(t)]}{\Exop\left[\rndx[\rndj] \mid \setH_1\right]}=0.\label{eq:unbiasednessunderh0}
\end{align}
Thus, it is left to show that \eqref{eq:unbiasedcondition} holds for $\setH=\setH_1$.

Under $\setH=\setH_1$, we have from \eqref{eq:crosscorr}
\begin{equation}
    \Exop\left[ \rndy(t)\rndx(t-\tau) \mid \setH_1 \right] = R_{\rnds\rnds}(\tau-\delta).
\end{equation}
Since $\rndy(\cdot)$ and $\rndx(\cdot)$ are zero-mean jointly Gaussian, the MMSE estimator of $\rndy(t_0)$ based on $\rndx(t_1)$ is given by
\begin{align}\label{eq:mmseforunbiasedness}
    \Exop\left[ \rndy(t_0) \mid \rndx(t_1) , \setH_1\right] = \rho(t_0-t_1)\cdot\frac{\sigma_{\rndy}}{\sigma_{\rndx}}\cdot\rndx(t_1).
\end{align}
Applying \eqref{eq:mmseforunbiasedness} with $t_0=t_{\rndj} + \tau$ and $t_1 = t_{\rndj}$ while conditioning on $\rndj$ (equivalently, on $t_{\rndj}$), we have
\begin{align}
    &\Exop\left[ \widehat{\rho}_{\lc}(\tau) \mid \setH_1 \right] \\
    &= \frac{\sigma_{\rndx}}{\sigma_{\rndy}}\cdot\frac{\Exop\left[ \rndy(t_{\rndj} + \tau) \mid \setH_1\right]}{\Exop\left[\rndx[\rndj]\mid \setH_1\right]} \\
    &= \frac{\sigma_{\rndx}}{\sigma_{\rndy}}\cdot\frac{\Exop\left[ \Exop\left[ \rndy(t_{\rndj} + \tau) \mid \setH_1, \rndx(t_{\rndj}), \rndj\right]  \mid \setH_1 \right]}{\Exop\left[\rndx[\rndj]\mid \setH_1\right]} \label{eq:lotetran1}\\
    &=\frac{\sigma_{\rndx}}{\sigma_{\rndy}}\cdot\frac{\Exop\left[ \Exop\left[ \rho(\tau)\cdot\frac{\sigma_{\rndy}}{\sigma_{\rndx}}\cdot\rndx(t_{\rndj})  \mid \setH_1,\rndx(t_{\rndj}) \right] \mid \setH_1 \right]}{\Exop\left[\rndx[\rndj]\mid \setH_1\right]} \\
    &=\rho(\tau)\cdot\frac{\Exop\left[ \rndx[\rndj]  \mid \setH_1 \right]}{\Exop\left[\rndx[\rndj]\mid \setH_1\right]} \label{eq:defofthecontmax}\\
    &= \rho(\tau),
\end{align}
where we have used the law of total expectation in \eqref{eq:lotetran1} and that, by definition, $\rndx(t_{\rndj})=\rndx[n]$ in  \eqref{eq:defofthecontmax}. Together with \eqref{eq:unbiasednessunderh0}, we have shown that $\Exop\left[ \widehat{\rho}_{\lc}(\tau) \mid \setH\right] = \rho(\tau)$, as desired.

\section{Closed-form expressions for $M_{\mathrm{in}}(j)$ and $M_{\mathrm{out}}(j)$}\label{app:explicittermscardinality}
Recall the definition \eqref{eq:defofintL}, as well as \eqref{eq:countableset} for any random $t_{\rndj} = \rndj T_{\mathrm{s}}$. For a fixed realization $\rndj=j$, this becomes $\Delta_{t_j}$.
Then every integer $n$ that satisfies $nT_{\mathrm{s}} \in [jT_{\mathrm{s}}-\delta_m,\;jT_{\mathrm{s}}+\delta_m]$ must obey $j-L \le n \le j+L$, and conversely any $n$ in this range satisfies $nT_{\mathrm{s}} \in [jT_{\mathrm{s}}-\delta_m,\;jT_{\mathrm{s}}+\delta_m]$. Hence
\begin{equation}\label{eq:appcardinalityD1}
    \Delta_{t_j} = \{\,j-L,\ldots,j+L\,\},
\end{equation}
so that the cardinality of the discrete search window is
\begin{equation}\label{eq:appcardinalityD2}
    D = \big|\Delta_{t_{\rndj}}\big| = 2L+1,
\end{equation}
which is \emph{independent} of $j$ (hence, deterministic) and depends on the continuous-time delay spread only through $L$ in \eqref{eq:defofintL}.

For a fixed $j\in\{0,\ldots,N-1\}$, the $D-1$ non-aligned indices in $\Delta_{t_j}$ can be written as
\begin{equation}
    \{n\in\Delta_{t_j}:n\neq j\}
    = \{\,j+\ell : \ell\in\{-L,\ldots,L\}\setminus\{0\}\,\}.
\end{equation}
An index $j+\ell$ is \emph{inside} the encoded block if $0\le j+\ell\le N-1$, and \emph{outside} otherwise. Thus, the number of inside non-aligned indices is
\begin{align}
    M_{\mathrm{in}}(j)
    &= \Big|\big\{\ell\in\{-L,\ldots,L\}\setminus\{0\} :
        0\le j+\ell\le N-1\big\}\Big|.
        \label{eq:Min-set-ct}
\end{align}
It is convenient to separate the left and right neighbors of $j$. The index $j$ has at most $L$ neighbors on each side inside the block, but fewer when $j$ is close to the boundaries. Specifically, define the sets of left and right non-aligned offsets by
\begin{align}
    \mathcal{L}(j)
    &\triangleq \big\{\ell\in\{-L,\ldots,-1\} : 0 \le j+\ell \le N-1\big\},\\
    \mathcal{R}(j)
    &\triangleq \big\{\ell\in\{1,\ldots,L\} : 0 \le j+\ell \le N-1\big\}.
\end{align}
Then the corresponding numbers of left and right neighbors are
\begin{equation}
    \big|\mathcal{L}(j)\big| = \min\{L,\,j\},
\end{equation}
and
\begin{equation}
    \big|\mathcal{R}(j)\big| = \min\{L,\,N-1-j\},
\end{equation}
respectively. Consequently, for all $0\le j\le N-1$,
\begin{align}
    M_{\mathrm{in}}(j)
    &= \big|\mathcal{L}(j)\big| + \big|\mathcal{R}(j)\big| \\
    &= \min\{L,\,j\} + \min\{L,\,N-1-j\}.
\end{align}
Since there are $D-1=2L$ non-aligned indices in total, the number of outside lags is
\begin{equation}
    M_{\mathrm{out}}(j)
    = (D-1) - M_{\mathrm{in}}(j)
    = 2L - M_{\mathrm{in}}(j).
    \label{eq:Mout-closed-ct}
\end{equation}

\bibliographystyle{IEEEbib}
\bibliography{./Inputs/refs}

\begin{thebibliography}{10}

\bibitem{li2002mimo}
Y.~G. Li, J.~H. Winters, and N.~R. Sollenberger,
\newblock ``{MIMO}-{OFDM} for wireless communications: Signal detection with
  enhanced channel estimation,''
\newblock {\em {IEEE} Trans. Commun.}, vol. 50, no. 9, pp. 1471--1477, 2002.

\bibitem{diao2022review}
P.~S. Diao, T. Alves, B. Poussot, and S. Azarian,
\newblock ``A review of radar detection fundamentals,''
\newblock {\em {IEEE} Trans. Aerosp. Electron. Syst.}, vol. 39, no. 9, pp.
  4--24, 2022.

\bibitem{abraham2002active}
D.~A. Abraham and P.~K. Willett,
\newblock ``Active sonar detection in shallow water using the page test,''
\newblock {\em {IEEE} J. Ocean. Eng.}, vol. 27, no. 1, pp. 35--46, 2002.

\bibitem{bachmann2012low}
C. Bachmann, M. Ashouei, V. Pop, M. Vidojkovic, H. De~Groot, and B. Gyselinckx,
\newblock ``Low-power wireless sensor nodes for ubiquitous long-term biomedical
  signal monitoring,''
\newblock {\em {IEEE} Commun. Mag.}, vol. 50, no. 1, pp. 20--27, 2012.

\bibitem{pasqualetti2013attack}
F. Pasqualetti, F. D{\"o}rfler, and F. Bullo,
\newblock ``Attack detection and identification in cyber-physical systems,''
\newblock {\em {IEEE} Trans. Autom. Control}, vol. 58, no. 11, pp. 2715--2729,
  2013.

\bibitem{chamberland2003decentralized}
J.-F. Chamberland and V.~V. Veeravalli,
\newblock ``Decentralized detection in sensor networks,''
\newblock {\em {IEEE} Trans. Signal Process.}, vol. 51, no. 2, pp. 407--416,
  2003.

\bibitem{chen2006channelaware}
B. Chen, L. Tong, and P. Varshney,
\newblock ``Channel-aware distributed detection in wireless sensor networks,''
\newblock {\em {IEEE} Signal Process. Mag.}, vol. 23, no. 4, pp. 16--26, 2006.

\bibitem{puccinelli2005wireless}
D. Puccinelli and M. Haenggi,
\newblock ``Wireless sensor networks: applications and challenges of ubiquitous
  sensing,''
\newblock {\em {IEEE} Circuits Syst. Mag.}, vol. 5, no. 3, pp. 19--31, 2005.

\bibitem{sarkar2014diat}
C. Sarkar, A.~U.~N. SN, R.~V. Prasad, A. Rahim, R. Neisse, and G. Baldini,
\newblock ``{DIAT}: A scalable distributed architecture for {I}o{T},''
\newblock {\em {IEEE} Internet Things J.}, vol. 2, no. 3, pp. 230--239, 2014.

\bibitem{qu2008cooperative}
Z. Qu, J. Wang, and R.~A. Hull,
\newblock ``Cooperative control of dynamical systems with application to
  autonomous vehicles,''
\newblock {\em {IEEE} Trans. Autom. Control}, vol. 53, no. 4, pp. 894--911,
  2008.

\bibitem{du2018sensable}
R. Du, P. Santi, M. Xiao, A.~V. Vasilakos, and C. Fischione,
\newblock ``The sensable city: A survey on the deployment and management for
  smart city monitoring,''
\newblock {\em {IEEE} Commun. Surveys Tuts.}, vol. 21, no. 2, pp. 1533--1560,
  2018.

\bibitem{tenney1981detection}
R.~R. Tenney and N.~R.~J. Sandell,
\newblock ``Detection with distributed sensors,''
\newblock {\em {IEEE} Trans. Aerosp. Electron. Syst.}, vol. AES-17, no. 4, pp.
  501--510, July 1981.

\bibitem{chair1986optimal}
Z. Chair and P.~K. Varshney,
\newblock ``Optimal data fusion in multiple sensor detection systems,''
\newblock {\em {IEEE} Trans. Aerosp. Electron. Syst.}, vol. AES-22, no. 1, pp.
  98--101, Jan. 1986.

\bibitem{viswanathan2002distributed}
R. Viswanathan and P.~K. Varshney,
\newblock ``Distributed detection with multiple sensors part {I}.
  fundamentals,''
\newblock {\em Proc. IEEE}, vol. 85, no. 1, pp. 54--63, 2002.

\bibitem{blum2002distributed}
R.~S. Blum, S.~A. Kassam, and H.~V. Poor,
\newblock ``Distributed detection with multiple sensors {II}. advanced
  topics,''
\newblock {\em Proc. IEEE}, vol. 85, no. 1, pp. 64--79, 2002.

\bibitem{varshney1996distributed}
P.~K. Varshney,
\newblock {\em Distributed Detection and Data Fusion},
\newblock Springer, Berlin, Germany, 1996.

\bibitem{ahlswede1986hypothesis}
R. Ahlswede and I. Csisz{\'a}r,
\newblock ``Hypothesis testing with communication constraints,''
\newblock {\em {IEEE} Trans. Inf. Theory}, vol. 32, no. 4, pp. 533--552, 1986.

\bibitem{han1987hypothesis}
T.~S. Han,
\newblock ``Hypothesis testing with multiterminal data compression,''
\newblock {\em {IEEE} Trans. Inf. Theory}, vol. 33, no. 6, pp. 759--772, 1987.

\bibitem{shimokawa1994error}
H. Shimokawa, T.~S. Han, and S. ichi Amari,
\newblock ``Error bound of hypothesis testing with data compression,''
\newblock in {\em Proc. IEEE Int. Symp. Inf. Theory (ISIT)}, Trondheim, Norway,
  1994, p. 114.

\bibitem{katz2016distributed}
G. Katz, P. Piantanida, and M. Debbah,
\newblock ``Distributed binary detection with lossy data compression,''
\newblock {\em {IEEE} Trans. Inf. Theory}, vol. 63, no. 8, pp. 5207--5227,
  2017.

\bibitem{sreekumar2018distributed}
S. Sreekumar and D. G{\"u}nd{\"u}z,
\newblock ``Distributed hypothesis testing over discrete memoryless channels,''
\newblock {\em {IEEE} Trans. Inf. Theory}, vol. 66, no. 4, pp. 2044--2066,
  2020.

\bibitem{sreekumar2019distributed}
S. Sreekumar and D. G{\"u}nd{\"u}z,
\newblock ``Distributed hypothesis testing over a noisy channel:
  Error-exponents trade-off,''
\newblock {\em Entropy}, vol. 25, no. 2, pp. 304, 2023.

\bibitem{carpi2021single}
F. Carpi, S.~E. Garg, and E. Erkip,
\newblock ``Single-shot compression for hypothesis testing,''
\newblock in {\em Proc. IEEE Int. Workshop Signal Process. Advances Wireless
  Commun. (SPAWC)}, 2021, pp. 176--180.

\bibitem{mao2024multi}
L. Mao, S. Yan, Z. Sui, and H. Li,
\newblock ``Multi-bit distributed detection of sparse stochastic signals over
  error-prone reporting channels,''
\newblock {\em {IEEE} Trans. Signal Inf. Process. Over Netw.}, vol. 10, pp.
  881--893, 2024.

\bibitem{kochman2025improved}
Y. Kochman and L. Wang,
\newblock ``Improved random-binning exponent for distributed hypothesis
  testing,''
\newblock {\em {IEEE} Trans. Inf. Theory}, 2025.

\bibitem{chan1978least}
Y. Chan, R. Hattin, and J. Plant,
\newblock ``The least squares estimation of time delay and its use in signal
  detection,''
\newblock {\em {IEEE} Trans. Acoust., Speech, Signal Process.}, vol. 26, no. 3,
  pp. 217--222, 1978.

\bibitem{weiss1983fundamentalpartI}
A. Weiss and E. Weinstein,
\newblock ``Fundamental limitations in passive time delay estimation--{P}art
  {I}: {N}arrow-band systems,''
\newblock {\em {IEEE} Trans. Acoust., Speech, Signal Process.}, vol. 31, no. 2,
  pp. 472--486, 1983.

\bibitem{weinstein1984fundamentalpartII}
E. Weinstein and A. Weiss,
\newblock ``Fundamental limitations in passive time-delay estimation--{P}art
  {II}: Wide-band systems,''
\newblock {\em {IEEE} Trans. Acoust., Speech, Signal Process.}, vol. 32, no. 5,
  pp. 1064--1078, 1984.

\bibitem{carter1987coherence}
G.~C. Carter,
\newblock ``Coherence and time delay estimation,''
\newblock {\em Proc. IEEE}, vol. 75, no. 2, pp. 236--255, Feb. 1987.

\bibitem{escamilla2020distributed}
P. Escamilla, M. Wigger, and A. Zaidi,
\newblock ``Distributed hypothesis testing: cooperation and concurrent
  detection,''
\newblock {\em {IEEE} Trans. Inf. Theory}, vol. 66, no. 12, pp. 7550--7564,
  2020.

\bibitem{inan2022fundamental}
Y. Inan, M. Kayaalp, A.~H. Sayed, and E. Telatar,
\newblock ``A fundamental limit of distributed hypothesis testing under
  memoryless quantization,''
\newblock in {\em Proc. IEEE Int. Conf. Commun. (ICC)}, 2022, pp. 4824--4829.

\bibitem{bounhar2024covert}
A. Bounhar, M. Sarkiss, and M. Wigger,
\newblock ``Covert distributed detection over discrete memoryless channels,''
\newblock in {\em Proc. IEEE Int. Symp. Inf. Theory (ISIT)}, 2024, pp.
  172--177.

\bibitem{bounhar2025dichotomy}
A. Bounhar, M. Sarkiss, and M. Wigger,
\newblock ``A dichotomy for distributed detection with limited communication,''
\newblock in {\em IEEE Inf. Theory Workshop (ITW)}, 2025.

\bibitem{fazel2012random}
F. Fazel, M. Fazel, and M. Stojanovic,
\newblock ``Random access sensor networks: Field reconstruction from incomplete
  data,''
\newblock in {\em Proc. {IEEE} Inf. Theory Appl. Workshop}, 2012, pp. 300--305.

\bibitem{sani2016distributed}
A. Sani and A. Vosoughi,
\newblock ``Distributed vector estimation for power-and bandwidth-constrained
  wireless sensor networks,''
\newblock {\em {IEEE} Trans. Signal Process.}, vol. 64, no. 15, pp. 3879--3894,
  2016.

\bibitem{li2020data}
G. Li and X. Song,
\newblock ``Data distribution optimization strategy in wireless sensor networks
  with edge computing,''
\newblock {\em IEEE Access}, vol. 8, pp. 214332--214345, 2020.

\bibitem{musluoglu2025distributed}
C.~A. Musluoglu and A. Bertrand,
\newblock ``Distributed blind source separation based on {F}ast{ICA},''
\newblock {\em {IEEE} Signal Process. Lett.}, 2025.

\bibitem{weiss2024joint}
A. Weiss, Y. Kochman, and G.~W. Wornell,
\newblock ``A joint data compression and time-delay estimation method for
  distributed systems via extremum encoding,''
\newblock in {\em Proc. IEEE Int. Conf. Acoust., Speech, Signal Process.
  (ICASSP)}, 2024, pp. 9366--9370.

\bibitem{weiss2025extremum}
A. Weiss, Y. Kochman, and G.~W. Wornell,
\newblock ``Extremum encoding for joint baseband signal compression and
  time-delay estimation for distributed systems,''
\newblock in {\em Proc. IEEE Int. Conf. Acoust., Speech, Signal Process.
  (ICASSP)}, 2025, pp. 1--5.

\bibitem{weiss2025joint}
A. Weiss, Y. Kochman, and G.~W. Wornell,
\newblock ``Joint data compression and time-delay estimation for distributed
  systems via extremum encoding,''
\newblock {\em {IEEE} Trans. Signal Process.}, 2025.

\bibitem{weiss2025asilomar}
A. Weiss and A. Lancho,
\newblock ``Extremum-based joint compression and detection for distributed
  sensing,''
\newblock in {\em Proc. Asilomar Conf. Signals, Syst., Comput.}, 2025.

\bibitem{zhang1988estimation}
Z. Zhang and T. Berger,
\newblock ``Estimation via compressed information,''
\newblock {\em {IEEE} Trans. Inf. Theory}, vol. 34, no. 2, pp. 198--211, 1988.

\bibitem{hadar2019distributed}
U. Hadar and O. Shayevitz,
\newblock ``Distributed estimation of {G}aussian correlations,''
\newblock {\em {IEEE} Trans. Inf. Theory}, vol. 65, no. 9, pp. 5323--5338,
  September 2019.

\bibitem{kochman2021communication}
Y. Kochman and L. Wang,
\newblock ``On the communication exponent of distributed testing for {G}aussian
  correlations,''
\newblock in {\em IEEE Inf. Theory Workshop (ITW)}, April 2021, pp. 1--5.

\bibitem{cai2024distributed}
T.~T. Cai and H. Wei,
\newblock ``Distributed {G}aussian mean estimation under communication
  constraints: Optimal rates and communication-efficient algorithms,''
\newblock {\em J. Mach. Learn. Res.}, vol. 25, no. 37, pp. 1--63, 2024.

\bibitem{HandbookExtremes2026}
M. {de Carvalho}, R. Huser, P. Naveau, and B.~J. Reich,
\newblock {\em Handbook of Statistics of Extremes},
\newblock Chapman \& Hall/CRC, Boca Raton, FL, 2026.

\bibitem{ito1963expected}
K. Ito,
\newblock ``The expected number of zeros of continuous stationary {G}aussian
  processes,''
\newblock {\em J. Math. Kyoto}, vol. 3, no. 2, pp. 207--216, 1963.

\bibitem{huang2019one}
X. Huang and B. Liao,
\newblock ``One-bit {MUSIC},''
\newblock {\em {IEEE} Signal Process. Lett.}, vol. 26, no. 7, pp. 961--965,
  April 2019.

\bibitem{bhandari2020one}
A. Bhandari, M.~H. Conde, and O. Loffeld,
\newblock ``One-bit time-resolved imaging,''
\newblock {\em {IEEE} Trans. Pattern Anal. Mach. Intell.}, vol. 42, no. 7, pp.
  1630--1641, April 2020.

\bibitem{weiss2021one}
A. Weiss and G.~W. Wornell,
\newblock ``One-bit direct position determination of narrowband {G}aussian
  signals,''
\newblock in {\em Proc. IEEE Workshop Stat. Signal Process. (SSP)}, July 2021,
  pp. 466--470.

\bibitem{eamaz2024uno}
A. Eamaz, K.~V. Mishra, F. Yeganegi, and M. Soltanalian,
\newblock ``{UNO}: Unlimited sampling meets one-bit quantization,''
\newblock {\em {IEEE} Trans. Signal Process.}, vol. 72, pp. 997--1014, 2024.

\bibitem{kumar2025carrier}
A. Kumar, K. Appaiah, and S.~R.~B. Pillai,
\newblock ``Carrier phase and frequency discriminators for receivers with
  one-bit quantization,''
\newblock {\em {IEEE} Trans. Commun.}, 2025.

\bibitem{fowler2005fisher}
M.~L. Fowler and M. Chen,
\newblock ``Fisher-information-based data compression for estimation using two
  sensors,''
\newblock {\em {IEEE} Trans. Aerosp. Electron. Syst.}, vol. 41, no. 3, pp.
  1131--1137, July 2005.

\bibitem{kolmogorov1956shannon}
A. Kolmogorov,
\newblock ``On the {S}hannon theory of information transmission in the case of
  continuous signals,''
\newblock {\em IRE Trans. Inform. Theory}, vol. {IT}-2, pp. 102--108, 1956.

\bibitem{ihara1993information}
S. Ihara,
\newblock {\em Information theory for continuous systems}, vol.~2,
\newblock World Scientific, 1993.

\bibitem{adler2007random}
R.~J. Adler and J.~E. Taylor,
\newblock {\em Random fields and geometry},
\newblock Springer, 2007.

\bibitem{piterbarg1996asymptotic}
V.~I. Piterbarg,
\newblock {\em Asymptotic methods in the theory of {G}aussian processes and
  fields}, vol. 148,
\newblock American Mathematical Soc., 1996.

\bibitem{azais2009level}
J.-M. Aza{\"\i}s and M. Wschebor,
\newblock {\em Level sets and extrema of random processes and fields},
\newblock John Wiley \& Sons, 2009.

\end{thebibliography}

\end{document}